\documentclass[a4paper,11pt]{article}
\pdfoutput=1 
\usepackage{jcappub} 
\usepackage{gensymb}
\usepackage{ulem}
\usepackage{qubic}
\usepackage{xcolor}
\usepackage{notoccite}
\usepackage[framemethod=tikz]{mdframed}
\definecolor{electricblue}{rgb}{0.49, 0.98, 1.0}
\newmdenv[
  backgroundcolor=electricblue,
  linewidth=0pt,
  topline=false,
  rightline=false,
  leftline=false]{myframe}{}
\usepackage{lipsum}

\usepackage{tcolorbox}

\title{QUBIC I: Overview and Science Program}
\author[1]{J.-Ch.~Hamilton}
\author[1]{L.~Mousset}
\author[2,3]{E.S.~Battistelli}
\author[1]{M.-A.~Bigot-Sazy}
\author[1]{P.~Chanial}
\author[1]{R.~Charlassier}
\author[2,3]{G.~D'Alessandro}
\author[2,3]{P.~de~Bernardis}
\author[2,3]{M.~De~Petris}
\author[4, 35]{M.M.~Gamboa Lerena}
% ...

\affiliation[1]{Universit\'e de Paris, CNRS, Astroparticule et Cosmologie, F-75006 Paris, France}
\affiliation[2]{Universit\a`{a} di Roma - La Sapienza, Italy}
\affiliation[3]{INFN sezione di Roma, 00185 Roma, Italy}
\affiliation[4]{Facultad de Ciencias Astron\a'{o}micas y Geof\a'{i}sicas (Universidad Nacional de La Plata), Argentina}
% ...
\affiliation[35]{CONICET, Argentina}

\abstract{

The Q \& U Bolometric Interferometer for Cosmology (QUBIC) is a novel kind of polarimeter optimized for the measurement of the B-mode polarization of the Cosmic Microwave Background (CMB), which is one of the major challenges of observational cosmology.  The signal is expected to be of the order of a few tens of nK, prone to instrumental systematic effects and polluted by various astrophysical foregrounds which can only be controlled through multichroic observations. QUBIC is designed to address these observational issues with a novel approach that combines the advantages of interferometry in terms of control of instrumental systematic effects with those of bolometric detectors in terms of wide-band, background-limited sensitivity. The QUBIC synthesized beam has a frequency-dependent shape that results in the ability to produce maps of the CMB polarization in multiple sub-bands within the two physical bands of the instrument (150 and 220 GHz). These features make QUBIC complementary to other instruments and makes it particularly well suited to characterize and remove Galactic foreground contamination. In this article, first of a series of eight, we give an overview of the QUBIC instrument design, the main results of the calibration campaign, and present the scientific program of QUBIC including not only the measurement of primordial B-modes, but also the measurement of Galactic foregrounds.  
We give forecasts for typical observations and measurements: with three years of integration on the sky and assuming perfect foreground removal as well as stable atmospheric conditions from our site in Argentina, our simulations show that we  can achieve a statistical sensitivity to the effective tensor-to-scalar ratio (including primordial and foreground B-modes)  $\sigma(r)=0.015$.

}

\date{\today}

\begin{document}

\maketitle

\section{Introduction}
\label{sec:intro}
A phase of exponential expansion called ``Inflation'' in the early Universe was proposed as a solution to major problems with the standard Big-Bang model: the horizon, flatness and monopole problems~\citep{1981PhRvD..23..347G,1982PhLB..108..389L}.  The horizon problem relates to the observed homogeneity of the Universe as evidenced by the smoothness of the Cosmic Microwave Background (CMB). The flatness problem is the fact that the observed curvature of
space-time is so close to flat in present times, that it must have been flat to enormous precision in the early universe.  The inflationary paradigm has been so successful in resolving the flatness and horizon problems, that it soon became a keystone component to the standard model of Cosmology. Observations of the CMB have provided several convincing arguments in favor of inflation: i) the series of acoustic peaks observed in the temperature power spectrum~\cite{debernardis_2000, netterfield_2002} ruled out the alternative to inflation, topological defects, as the mechanism responsible for the dominant primordial density fluctuations ; ii) the observation of correlations between temperature and E-polarization of the CMB~\cite{kogut_2003} as well as the observed phase opposition between peaks in the $TT$ and $EE$ power spectra~\cite{2004Sci...306..836R, 2009ApJ...692.1247P} are evidence for adiabatic primordial perturbations, such as produced by inflation ; iii) the measurement of the CMB scalar spectral index slightly lower than one~\cite{2020A&A...641A...6P} is a prediction from inflation. However, despite these strong arguments, all showing agreement between observations and inflation, there continues to be a lack of direct observational evidence of inflation despite the 40~years that have passed since it was first proposed.

One observable effect of Inflation is the generation of polarization B-modes in the CMB. This happens because during the exponential expansion of the Universe intense gravitational waves are generated.  The gravitational waves induce re-orientation of the primordial plasma such that B-modes are visible in the radiation released at the surface of last scattering, which is the CMB.  This weak signal has not been detected so far and  many projects are operating, or are proposed, to measure the polarization B-modes in the CMB. These projects include SPTPol~\cite{sayre2019measurements}, POLARBEAR~\cite{ade2014measurement}, ACTPol~\cite{louis2017atacama}, BICEP2~\cite{ade2018constraints}, CLASS~\cite{dahal2020class}, POLARBEAR 2 + Simons Array~\cite{Polarbear+simons2016}, advanced ACT~\cite{AdvACT2016}, upgrade of the BICEP3/Keck array~\cite{grayson2016bicep3}. Planned experiments include Simons Observatory~\cite{2019JCAP...02..056A},  PIPER~\cite{piper2016}, LSPE~\cite{addamo2020large}, CMB-S4~\cite{cmbs4}, LiteBIRD \cite{ hazumi2019} and PICO~\cite{PICO}.

The use of bolometric interferometry to measure polarization B-modes in the CMB was proposed for a series of projects leading up to QUBIC. These are the Millimeter-wave Bolometric Interferometry~\citep[MBI,][]{2002AIPC..616..126A,2003NewAR..47.1173T}, the Einstein Polarization Interferometer for Cosmology~\citep[EPIC,][]{2006NewAR..50..999T}, the Background RAdiation INterferometer \citep[BRAIN,][]{2007NewAR..51..256P}, and CMBPol\citep{2009JPhCS.155a2003T}. Members of all these collaborations joined together to develop the \mbox{Q \& U Bolometric} Interferometer for Cosmology \citep[QUBIC,][]{2009arXiv0910.0391K,2009AIPC.1185..506G,2011APh....34..705Q,2012MNRAS.423.1293B,2016arXiv160904372A}.

This paper is part of a special issue on QUBIC which includes details on all its design aspects, as well as on the performance of an advanced prototype, the Technological Demonstrator (TD).  The scientific overview and expected performance of the instrument are addressed here.  The other papers in the special issue address the following topics: the ability of bolometric interferometry to do spectral imaging~\citep{2020.QUBIC.PAPER2}, the calibration and performance in the laboratory of the TD~\citep{2020.QUBIC.PAPER3}, the performance of the detector array and readout electronics~\citep{2020.QUBIC.PAPER4}, the cryogenic system performance~\citep{2020.QUBIC.PAPER5}, the Half Wave Plate rotator system~\citep{2020.QUBIC.PAPER6}, the back-to-back feedhorn-switch system~\citep{2020.QUBIC.PAPER7}, and the optical design and performance~\citep{2020.QUBIC.PAPER8}.

This paper is organized as follows.  A description of bolometric interferometry is provided in Section~\ref{sec:boloint}. This is followed in Section~\ref{sec:sciobj} with an overview of the scientific objectives, the current state-of-the-art of CMB polarization experiments, and the expected performance of QUBIC. Finally, conclusions are presented in Section~\ref{sec:conclusions}.

\section{Bolometric interferometry and QUBIC}
\label{sec:boloint}
\subsection{Bolometric interferometry}
\label{subsec:boloint}

The idea of bolometric interferometry dates back to the 19$^\mathrm{th}$ century when Fizeau proposed an adding interferometer to measure the expected difference in the velocity of light traveling in moving media \citep{1851fizeau}. The setup split the signal from the Sun into two pipes, each filled with water flowing in  opposite directions. The beams were then recombined and the interference fringes measured to determine the phase difference between them. In QUBIC, the pipes are replaced by an array of back-to-back feedhorns, and the beams are recombined by an arrangement of two mirrors \citep[see O'Sullivan et al.][for details]{2020.QUBIC.PAPER8}.

An important advantage of the Fizeau arrangement is its inherent wide band performance. There is no wavelength selectivity which results in an ideal performance at any particular wavelength. This can be compared to the case for example of the Martin-Puplett interferometer \citep{1970InfPh..10..105M} which introduces a quarter wavelength optical path difference by moving one of the corner mirrors, making it tuned to a particular wavelength.

Imaging radio interferometers generally work with electronic signals where multiplicative correlation is a natural operation with electronic components, both analog and digital.  The correlations between signals from different antennas build up a sampling of the ``\uvplane''. The Fourier transform of the \uvplane\ is an image of the sky, usually called the ``dirty image'' because no compensation has been done for the under-sampling of the \uvplane, and no calibration has been done \citep[see for example Thompson, Moran and Swenson][]{2017isra.book.....T}. In a bolometric interferometer, this ``dirty image'' is formed directly by the optical combiner, and is recorded by the detector array.  QUBIC treats the entire band ``in one go'', whereas a traditional radio interferometer using narrow band electronic processors must split the band into many sub-bands, making large bandwidths expensive to realize in terms of processing and power requirements.

\begin{figure}[t]
\centering
\includegraphics[ width = 1 \hsize ]{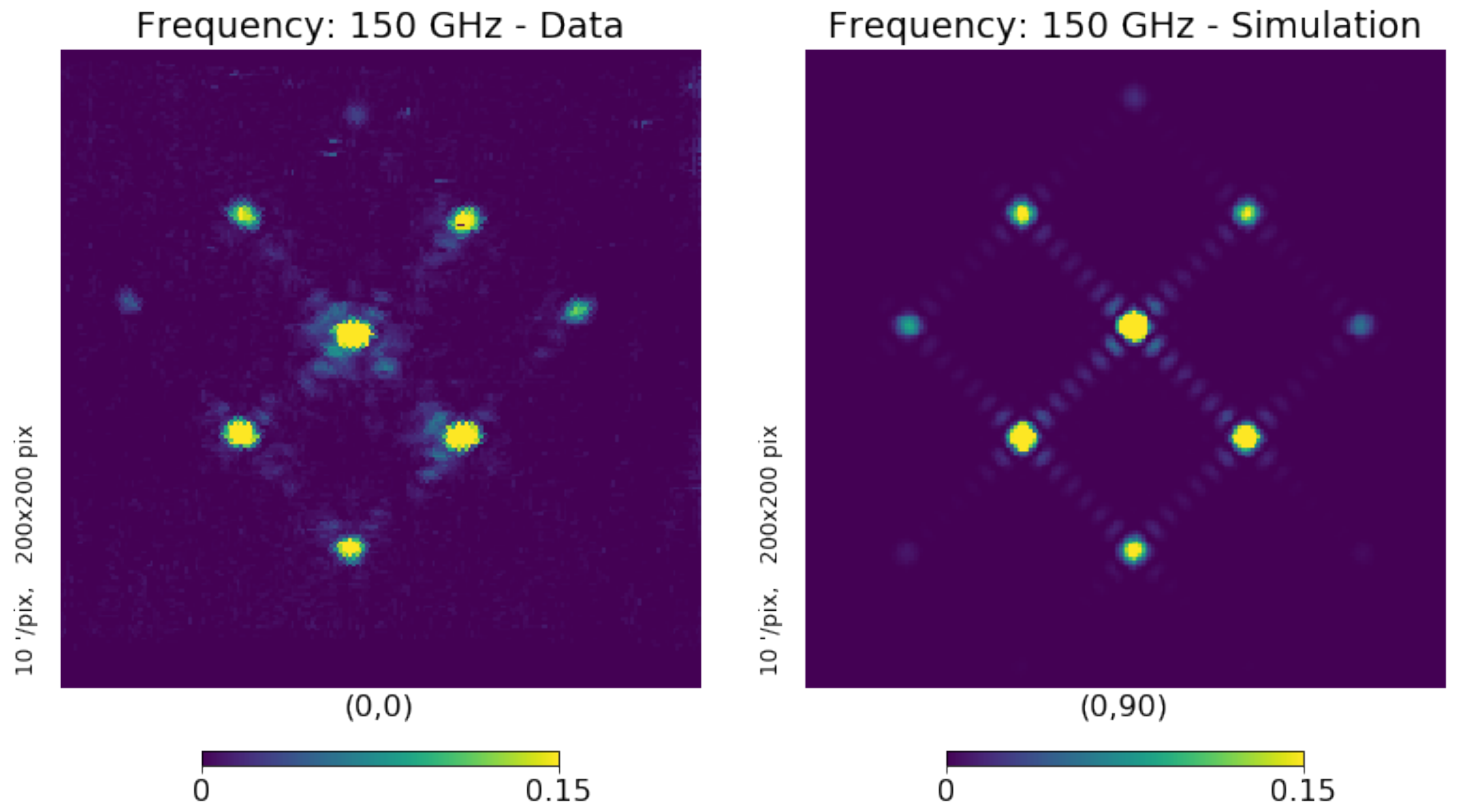}
\caption{QUBIC synthesized beam on the sky (left: laboratory measurement with the TD, right: simulations without optical aberrations) at 150~GHz. Note that the color scales are arbitrary units. Details on this measurement can be found in Torchinsky et al.~\cite{2020.QUBIC.PAPER3} from this series of articles. The synthesized beam shrinks with increasing frequency as can be seen with the animated version of this image that can be found online at \href{https://box.in2p3.fr/index.php/s/bzPYfmtjQW4wCGj}{https://box.in2p3.fr/index.php/s/bzPYfmtjQW4wCGj}.}
\label{Fig:sbqubic}
\end{figure}

This method is equivalent to imaging the sky with an imager whose beam is the synthesized beam of the bolometric interferometer (the ``dirty beam'' of the interferometer), formed by the combination of all interference patterns of all possible pairs of horns in the aperture array~\cite{2011APh....34..705Q}. In the case of the QUBIC Full Instrument, 400~horns form the synthesized beam shown in figure~\ref{Fig:sbqubic} for one specific detector. This is the beam the detector located in this position in the focal plane scans the sky with. Note that the synthesized beam differs for different detectors in the focal plane (see~Mousset et al.~\cite{2020.QUBIC.PAPER2} and O'Sullivan et al.~\cite{ 2020.QUBIC.PAPER8} from this series of articles for details). The synthesized beam is very different from a classical imager beam as it exhibits multiple peaks (because the horn array has a finite extension), the angular distance between two peaks is driven by the ratio between the wavelength and the distance between two nearby horns while the angular resolution of each peak is driven by the ratio between the wavelength and the maximal distance between horns. This is detailed in O'Sullivan~et.~al~\cite{2020.QUBIC.PAPER8} from this series of articles. In the case of QUBIC, with 400~horns, the angular resolution achieved is 23.5~arcminutes at 150~GHz.
Being the result of the summation of interference fringes, this synthesized beam significantly evolves with electromagnetic frequency, which is the basis of the spectral imaging capabilities of QUBIC (described briefly in section~\ref{sec:specimg} and in detail in~Mousset et al.~\cite{2020.QUBIC.PAPER2} in this series of articles).

\subsection{Self-Calibration}
\label{sec:selfcal}

Self calibration in aperture synthesis evolved from the idea of ``phase-closure''~\cite{1974A&AS...15..417H,1976A&A....50...19F,1978ApJ...223...25R} in a phased-array radiotelescope. Signals from the individual antenna elements of a phased-array are combined together and phase differences between the signals are corrected such that the combination of all the signals is equivalent to the signal captured by a large single-dish antenna pointing to the desired direction. With the advent of digital correlators, the correlation coefficients could be saved and a more sophisticated processing resulted in improved calibration. This came to be known as ``self-calibration''~\cite{1981MNRAS.196.1067C,1984ARA&A..22...97P}.

Bolometric interferometry can also take advantage of the self-calibration used in traditional aperture synthesis in radio astronomy. 
However, as a bolometric interferometer directly observes the dirty image, one needs to `go backwards'' from the dirty image and reconstruct the \uvplane.
The detailed procedure and analysis technique are described in Bigot-Sazy~et~al.~\cite{2013A&A...550A..59B}.
The general idea is to observe a polarized  artificial point source with the bolometric interferometer with only one pair of horns open at a time while the others are closed\footnote{In practice, in order to keep a roughly constant loading on the detector array, we achieve this observation through measuring successively the point source with one horn closed, all the others being open, then only the second horn closed, then both closed and finally all horns open. We eventually combine these observations to achieve the same measurement as with a single pair of horns open.}, which can be achieved using the mechanical shutters (also called RF switches) developed for QUBIC that can open or close each horn (see Cavaliere~et~al.~\cite{2020.QUBIC.PAPER7} in this series of articles for details on the QUBIC horns and RF switches). 
In the absence of instrumental systematic effects, two ``redundant'' pairs of horns (same distance and orientation) would correspond to the exact same visibility (Fourier mode on the sky) and the bolometric interferometer would therefore measure the same quantity. A difference between the two can only come from instrumental effects\footnote{Small-scale atmospheric fluctuations could also be responsible for a difference between measurements performed with ``redundant'' pairs of horns. However, we anticipate this to be a marginal effect because i) the artificial source has a strong emission (a fraction of Watt) and ii) self-calibration will be  performed scanning the source over long periods, averaging-out the atmospheric fluctuations.}.
By measuring in this manner all possible visibilities, one can fit parameters of a very general instrument model comprising hundreds of parameters (a few for each horn, bolometer) as well as for different electromagnetic frequencies if the source can be tuned in frequency. The number of constraints scales as the number of horns squared while the number of unknowns is proportional to the number of horns. As a result the problem is heavily over constrained for an instrument like QUBIC. The following instrumental systematics, expected to be the major ones for a bolometric interferometer, have been considered~\cite{2013A&A...550A..59B}:
\begin{itemize}
    \item uncertainty on the exact location of the horns which would affect the shape of the synthesized beam;
    \item uncertainty on the exact location of the bolometers in the focal plane. Similarly, it would affect the shape of the synthesized beam because it gradually changes across the focal plane (see O'Sullivan et al.~\cite{2020.QUBIC.PAPER8});
    \item relative gains of the bolometers;
    \item uncertainties on the primary beam (towards the sky) shape;
    \item uncertainties on the secondary beam (towards the detectors) shape;
    \item imperfections of the successive optical elements, each modeled by a Jones matrix containing differential gains for each polarization orientation as well as cross-polarization.
\end{itemize}
We have shown in Bigot-Sazy et al.~\cite{2013A&A...550A..59B} that spending 1 second on self-calibration for each baseline would reduce the $E$ to $B$ power spectrum leakage by an order of magnitude for typical values for the above systematic effects. The leakage drops to two orders of magnitude with 100 seconds per baseline.

Self-calibration is a special feature of bolometric interferometry for measuring instrumental systematic effects and subsequently accounting for them in the data analysis process through a more accurate modeling of the synthesized beam than the purely theoretical one, hence improving the performance of the map-making (see section~\ref{sims}).

\subsection{Spectral Imaging}
\label{sec:specimg}
\begin{figure}[t]
\centering
\includegraphics[ width =\hsize ]{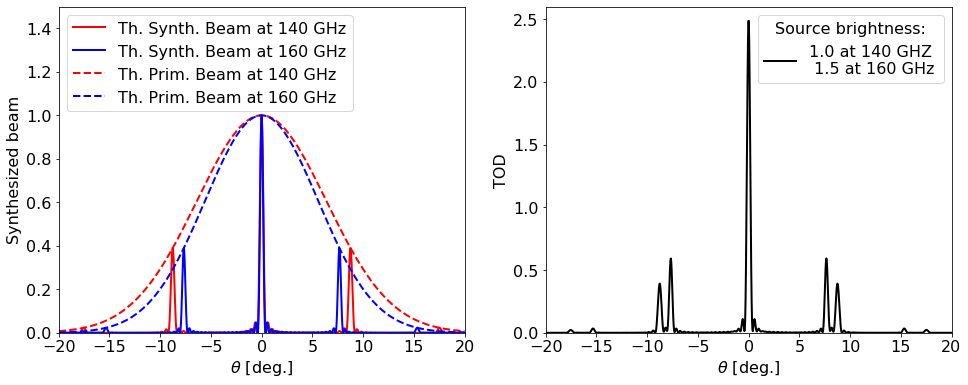}
\caption{{\bf (left)} A cut of the QUBIC theoretical synthesized and primary beams at two different frequencies for a bolometer located at the center of the focal plane. The figure shows how the location of the multiple peaks significantly changes for a small frequency change. This behaviour is at the basis of the spectral imaging capabilities offered by bolometric interferometry. {\bf (left)} signal detected by the same bolometer when scanning across an imaginary point source emitting 1 and 1.5 at 140 and 160 GHz respectively: we have both spatial and frequency information.}
\label{Fig:sb_freq}
\end{figure}

Spectral imaging is described in detail in Mousset~et~al.~\cite{2020.QUBIC.PAPER2} from this series of articles, but we provide here a short explanation of this specific feature offered by bolometric interferometry.
As mentioned in section~\ref{subsec:boloint} and shown in figure~\ref{Fig:sbqubic}, the multiply peaked shape of the synthesized beams evolves with frequency. A cut of the QUBIC synthesized beam is shown in figure~\ref{Fig:sb_freq} (left) for two different frequencies. 
As the respective distance between peaks changes with frequency, the signal detected with a given bolometer will combine (through convolution with the synthesized beam) different directions on the sky for different incoming frequencies within the physical wide band. Let's imagine, as a simple toy model, that the sky consists in a single point source emitting only at two distinct frequencies 140 and 160 GHz with relative brightness 1 and 1.5. As the instrument scans over the point source, the response of the Time-Ordered-Data (TOD) will be the sum of the synthesized beam at 140 and 160 GHz scaled by 1 and 1.5 respectively. This is shown in figure~\ref{Fig:sb_freq} (right) as a function of the angle from the point source. We can easily reconstruct the location of the source from the position of the central peak, as well as the spectrum of the source from the amplitudes of each of the peaks around the central one.
These signals from different frequencies will only be significantly different from each other if the corresponding side-peaks are more separated than their intrinsic width. 

This shows how one can simultaneously recover spatial and frequency information within a wide physical band for a bolometric interferometer. Of course, the above toy-model corresponds to a source that is not extended in spatial nor frequency domain, but we have shown in Mousset et al.~\cite{2020.QUBIC.PAPER2} that the exact same reasoning can be applied to diffuse signals with continuous Spectral Energy Distribution (SED).
This spectral imaging reconstruction is done at the map-making stage in the data-analysis pipeline with data collected within a wide bandwidth. It can be done by reconstructing maps in as many different sub-frequencies as allowed by the ratio between frequency shift and intrinsic peak width in the synthesized beam. We have shown in Mousset et al.~\cite{2020.QUBIC.PAPER2} that the corresponding relative frequency resolution FWHM is $\Delta_\nu/\nu \sim 1/P$ where $P$ is the maximum number of apertures along an axis of the bolometric interferometer horn array. For a 20x20 horn-array, QUBIC therefore achieves 5 sub-bands with a frequency FWHM resolution of $\Delta_\nu/\nu\sim 0.05$ within each of the 25\% bandwidth frequency bands.

In the current context of search for primordial B-modes in the CMB originating from tensor perturbations from inflation, spectral imaging presents an opportunity to constrain foregrounds alternative to other current approaches. Previous analyses have indeed shown that B-modes measurements are largely dominated by foregrounds~\cite{2014PhRvL.112x1101B,2015A&A...576A.104P,2015PhRvL.114j1301B,2018PhRvL.121v1301B} which can only be removed through their frequency behaviour which  is distinct from that of the CMB. Classical imagers usually approach this issue by multiplying the number of frequencies at which they observe the CMB. Galactic dust is currently the most worrying foreground at frequencies above 100\,GHz. It can be constrained from the ground through wide atmospheric windows around 150 and 220~GHz. The noise severely increases at higher frequencies because of atmospheric emissivity. As a result, constraints on foregrounds can only be achieved through comparisons between these few, largely separated frequency bands. While their large separation in frequency may appear as an advantage as it increases lever arm, it is also a limitation as it prevents data analyses to consider realistic electromagnetic spectra for dust emission that could exhibit changes of slope or dust decorrelation between frequencies~\cite{2017A&A...599A..51P, 2018PhRvL.121v1301B}. With spectral imaging, one can measure the spectrum of the foreground components {\bf locally} (i.e., within the bandwidth) and avoid large extrapolations between distant frequencies. This local approach to foreground mitigation is complementary to the usual approach with largely separated frequency bands. This is studied in section~\ref{sect:foreground_challenge} of this article.

\subsection{The QUBIC Instrument}
In order to achieve bolometric interferometry, QUBIC relies on an optical system consisting of back-to-back horns that select the relevant baselines and an optical combiner focusing on a bolometric focal plane. The optical combiner forms interference fringes while the bolometers average their powers over timescales much larger than the period of the electromagnetic light. This is therefore the optical equivalent of a (wide-band) correlator in classical interferometry. Being a bolometric device, the whole instrument operates at cryogenic temperatures thanks to a large cryostat described in Masi et al.~\cite{2020.QUBIC.PAPER5} from this series of articles.

\begin{figure}[p!]
\centering
$\vcenter{\hbox{\includegraphics[ width = 0.9 \hsize ]{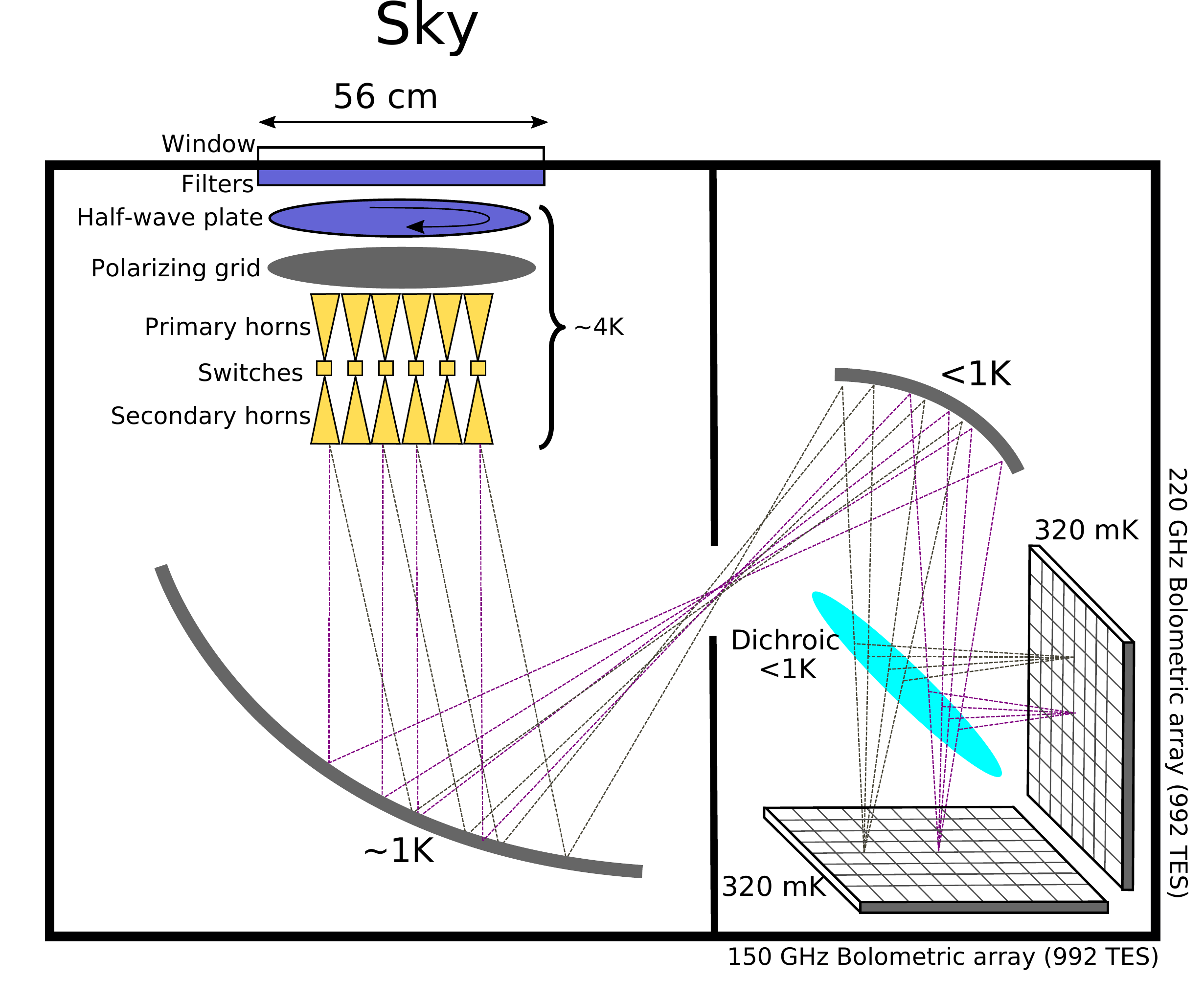}}}$
$\vcenter{\hbox{\includegraphics[ width = 0.7 \hsize ]{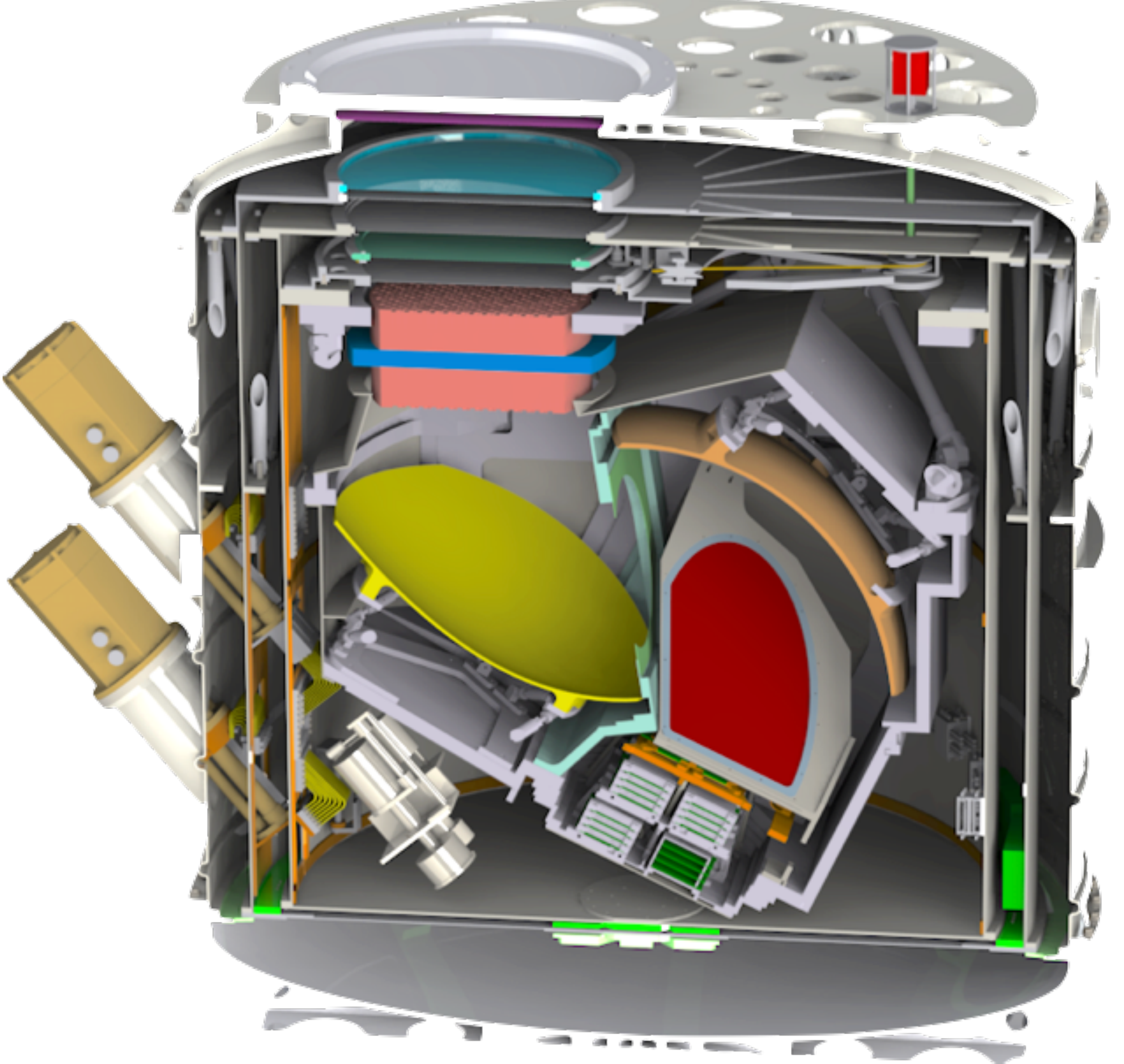}}}$
\caption{Schematic of the QUBIC instrument \textbf{(top)} and sectional cut of the cryostat \textbf{(bottom)} showing the same sub-systems in their real configuration.}
\label{Fig:qubic_scheme}
\end{figure}

\begin{table}[t]
    \renewcommand{\arraystretch}{1.}
    \begin{center}
        \caption{\label{tab_qubic_params}QUBIC main parameters}
        \begin{tabular}{p{6.4cm} m{2.5cm} m{5.5cm}}
            \hline
            Parameter& TD & FI \\
            \hline
            \hline
            {\bf Instrument}\\
            \hline
            Frequency channels \dotfill &150 GHz & 150 GHz \& 220 GHz\\
            Frequency range 150 GHz \dotfill &[131-169] GHz &[131-169] GHz\\
            Frequency range 220 GHz \dotfill &- &[192.5-247.5] GHz\\
            Window Aperture [m]\dotfill & 0.56 & 0.56 \\
            Number of horns\dotfill &64 &400\\
            Number of detectors\dotfill &248 &992$\times$2\\
            Detector noise [$\mathrm{W/\sqrt{Hz}}$]\dotfill & 2.05$\times 10^{-16}$ & 4.7$\times 10^{-17}$ \\
            Focal plane temp. [mK]\dotfill &300 &300\\
            Synthesized beam FWHM [deg]\dotfill &0.68~\cite{2020.QUBIC.PAPER3} &0.39 (150 GHz), 0.27 (220 GHz)\\
            \hline
            \hline
            {\bf Scanning Strategy (see sect.~\ref{sims})}\\
            \hline
            Elevation range [deg]\dotfill &[30-70] &[30-70]\\
            Azimuth scan width [deg]\dotfill &$\pm 17.5$ &$\pm 17.5$\\
            Azimuth scan speed [deg/s]\dotfill &0.4 &0.4\\
            Sky Coverage\dotfill &1.5\% &1.5\%
        \end{tabular}
    \end{center}
\end{table}

A schematic of the design of QUBIC is shown in figure~\ref{Fig:qubic_scheme} and the main instrument parameters are listed in table~\ref{tab_qubic_params}. The sky signal first goes through a 56~cm diameter window made of Ultra-High-Molecular-Weight Polyethylene followed by a series of filters cutting off frequencies higher than the desired ones. The next optical component is a stepped rotating Half-Wave-Plate which modulates incoming polarization. This sub-system is described in D'Alessandro~et~al.~\cite{2020.QUBIC.PAPER6} from this series of articles. A single polarization is then selected thanks to a polarizing grid. Although reflecting half of the incoming photons may appear as a regrettable loss, it is in fact one of the key features of QUBIC for handling instrumental systematics, especially polarization-related ones: a single polarization is selected just after polarization modulation by a wire-grid, while our bolometers are not sensitive to polarization due to the XY symmetry of the absorbing grid (see Piat et al~\cite{2020.QUBIC.PAPER4}). As a result, any cross-polarization occurring after the polarizing grid (horns, uncontrolled reflections inside the optical combiner) are negligible. This has been confirmed during calibration by measuring the cross-polarization of our TES (see figure~\ref{Fig:fringe}-right where we show a cross-polarization measurement below 0.4\% at 95\% C.L. for a TES). The next optical device is an array of 400 back-to-back corrugated horns made of an assembly of two 400-horns arrays, composed of 175 aluminium platelets (0.3 mm thick) chemically etched to reproduce the corrugations required for the horns to achieve the required performance. An array of mechanical shutters (RF switches) separates the two back-to-back horn arrays in order to be able to close or open horns for self-calibration (see section~\ref{sec:selfcal}). Both front and back horns are identical, each with a field of view of 13~degrees FWHM with secondary lobes below $-$25~dB. The horns and switches are described in detail in Cavaliere~et. al~\cite{2020.QUBIC.PAPER7} from this series of articles. The back-horns directly illuminate the two-mirrors off-axis Gregorian optical combiner (described in detail in O'Sullivan et al.~\cite{2020.QUBIC.PAPER8}) that focuses the signal onto the two perpendicular focal planes, separated by a dichroic filter that splits the incoming waves into two wide bands centered at 150~GHz for the on-axis focal plane and 220~GHz for the  off-axis one. The focal planes are each equipped with 992~NbSi Transition-Edge-Sensors (the detection chain is described in detail in Piat et al.~\cite{2020.QUBIC.PAPER4} from this series of articles) cooled down to 300~mK using a sorption fridge. A realistic view of the cryostat can be seen in the right panel of figure~\ref{Fig:qubic_scheme}. The cryostat weights roughly 800~kg and is around 1.6~meter high for a 1.4~meter diameter.

\begin{figure}[t]
\centering
\includegraphics[ width = 0.64 \hsize ]{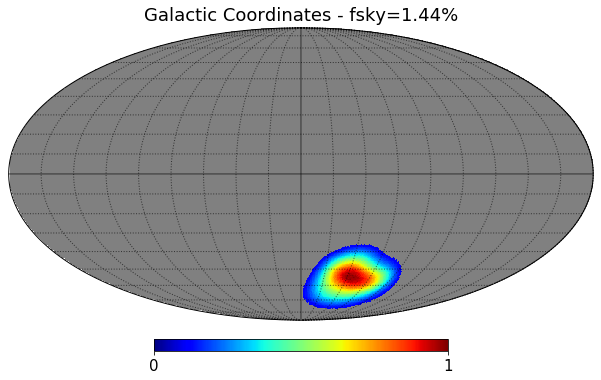}\includegraphics[ width = 0.36 \hsize ]{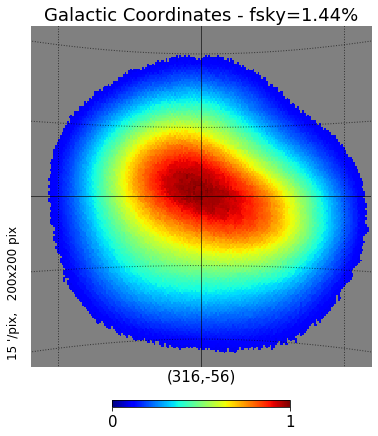}
\caption{Baseline sky coverage in Galactic coordinates (see table~\ref{tab_qubic_params} for corresponding scanning strategy parameters). The coverage is cut at 10\% from the maximum and achieves a sky fraction of 1.5\%. The grid spacing in latitude and longitude are 10 and 20 degrees respectively.}
\label{Fig:coverage}
\end{figure}

The anticipated scanning strategy for QUBIC is back and forth azimuth scans at constant elevation, following the azimuth of the center of the field as a function of sidereal time and updating elevation regularly in order to have scans with as many angles as possible in sky coordinates. The latitude of the QUBIC site allows us to cover a wide range of angles. The detailed scanning strategy will be determined when observing the real sky based on optimal mitigation of atmospheric fluctuations. We show in figure~\ref{Fig:coverage} a typical sky coverage obtained with the baseline scanning parameters given in table~\ref{tab_qubic_params}.

\subsubsection{The QUBIC Technological Demonstrator}
\label{sec_qubic_td}
\begin{figure}[t]
\centering
\includegraphics[ width = 0.36 \hsize ]{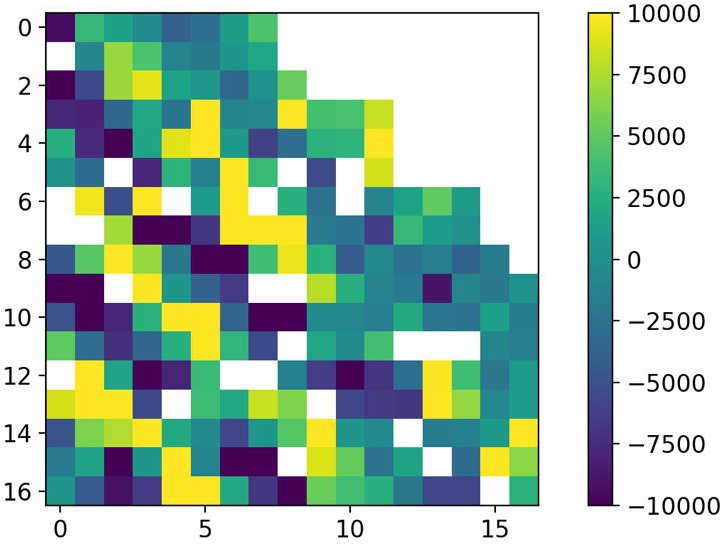}\includegraphics[ width = 0.36 \hsize ]{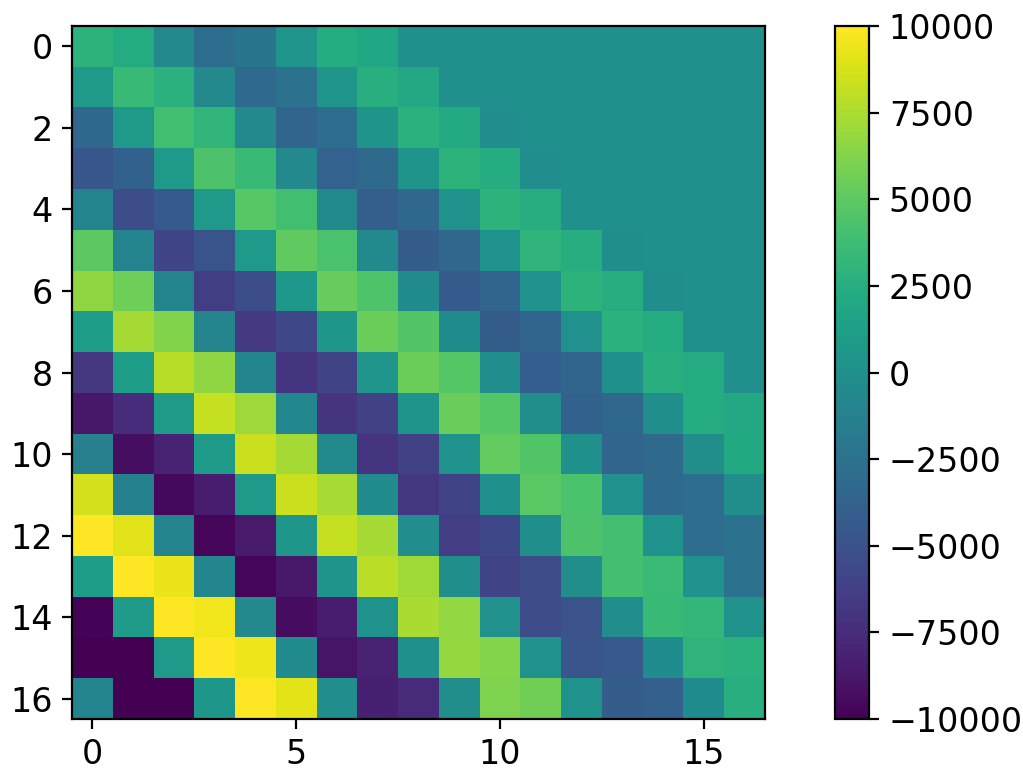}\includegraphics[ width = 0.28\hsize ]{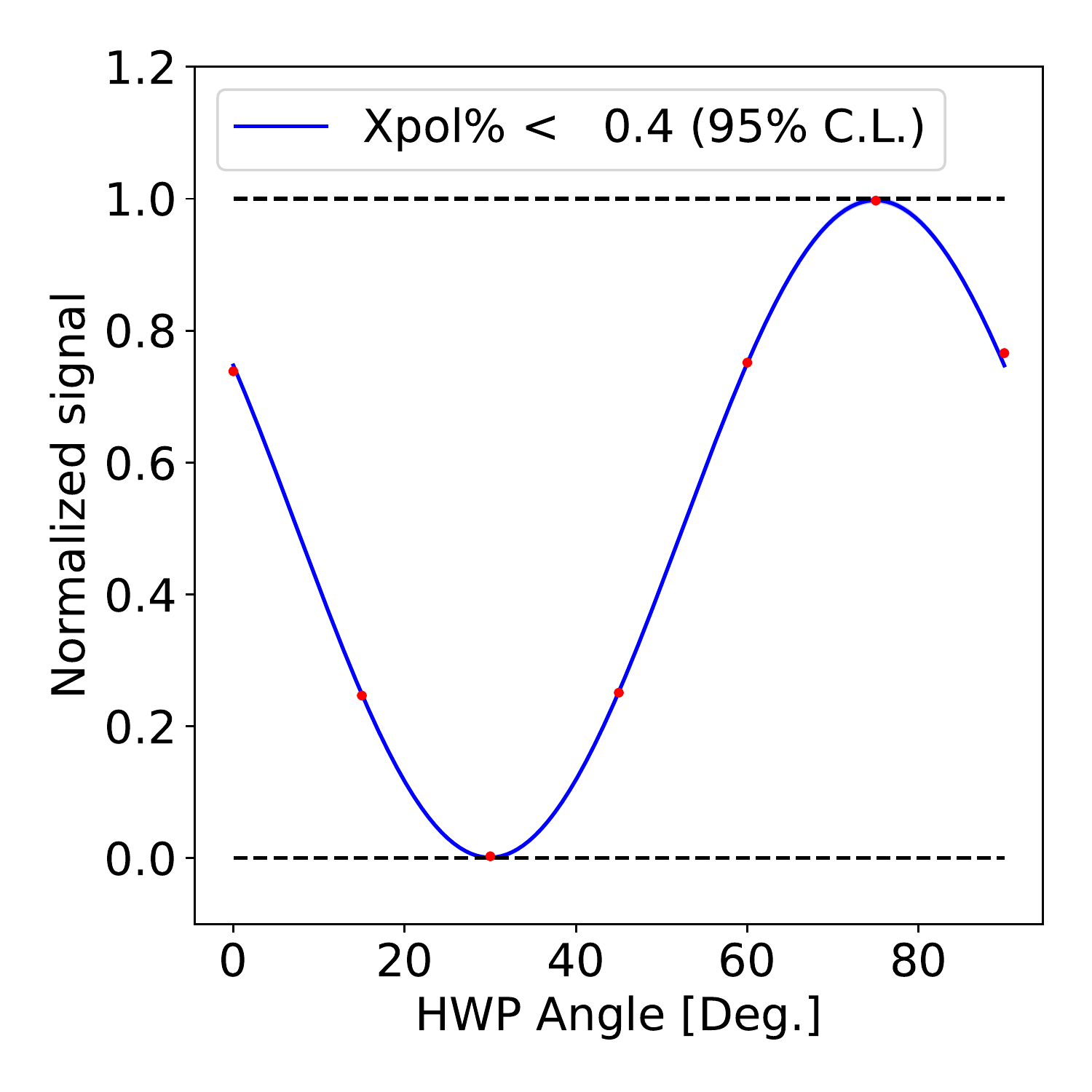}
\caption{{\bf (left and center)} Fringe pattern measured (left) and simulated (center) with the QUBIC TD  (coordinates are in the on-axis focal plane and units arbitrary) using the mechanical shutters~\cite{2020.QUBIC.PAPER7} in order to achieve a measurement equivalent to that of a single baseline (two horns open, all others closed). Units are arbitrary. {\bf (right)} Cross-Polarization measured by rotating the Half-Wave-Plate and observing our polarized calibration source. This result is obtained with a TES bolometer close to the focal plane center and showing a high signal-to-noise ratio.}
\label{Fig:fringe}
\end{figure}

The QUBIC Technological Demonstrator (hereafter QUBIC TD) uses the same cryostat, cooling system, filters and general sub-system architecture as described above but with only 64~back-to-back horns and mirrors reduced according to the illumination of the 64~horns. It also uses a single 248~TES bolometer array operating at 150~GHz. The QUBIC TD has been used as an intermediate step before the Full Instrument (FI) in order to characterize and demonstrate bolometric interferometry in the laboratory. We have confirmed the expected behaviour of the instrument and all the anticipated specific features of bolometric interferometry during this calibration campaign which is detailed in Torchinsky~et~al.~\cite{2020.QUBIC.PAPER3}. A selection of the most relevant results from the calibration includes:
\begin{itemize}
    \item the measurement of the bolometric interferometer synthesized beams multiple peaked shape in overall agreement with the theoretical prediction we made in Battistelli et al.~\cite{2011APh....34..705Q}, shown in figure~\ref{Fig:sb_freq}. The difference between theory and measurement in the locations and amplitudes of the peaks is expected from optical aberrations (see~\cite{2020.QUBIC.PAPER8}) and can be calibrated from the data for the map reconstruction;
    \item the evolution as a function of frequency of the inter-peak separation in the synthesized beam (available online at \href{https://box.in2p3.fr/index.php/s/bzPYfmtjQW4wCGj}{https://box.in2p3.fr/index.php/s/bzPYfmtjQW4wCGj}) which is at the basis of Spectral Imaging described briefly in section~\ref{sec:specimg} and in detail in Mousset et al.~\cite{2020.QUBIC.PAPER2}. This possibility is studied in detail in section~\ref{sect:foreground_challenge} of this article;
    \item the measurement of individual fringe patterns using the mechanical shutters~\cite{2020.QUBIC.PAPER7} shown in the left panel of figure~\ref{Fig:fringe} in this article and a more detailed description in figure~13 of Torchinsky et al.~\cite{2020.QUBIC.PAPER3}. As discussed in section~\ref{sec:selfcal}, this measurement is at the basis of the self-calibration allowing QUBIC to achieve a novel control of instrumental systematic effects, as shown in Bigot-Sazy et al.~\cite{2013A&A...550A..59B};
    \item Figure~\ref{Fig:fringe} shows a measurement of the TD cross-polarization with an upper-limit 0.4\% at 95\% C.L. for a detector close to the center of the focal plane. As studied in details in D'Alessandro et al.~\cite{2020.QUBIC.PAPER6} from this series of articles, we have found a median measured cross-polarization of 0.12\% among our detectors and a 0.61\% median 95\% upper-limit.
    77\% of our detectors have a measured cross-polarization compatible with zero at the one-sigma level. Such low cross-polarization is an important feature for detecting a signal as small as the primordial B-modes. This is achieved in QUBIC thanks to its specific polarization design with a single polarization selected before the interferometer apertures and full-power detectors as discussed above in this section.
    \item The intrinsic detector noise has been measured to be 4.7$\times 10^{-17}~\mathrm{W/\sqrt{Hz}}$ but due to noise aliasing with the TD readout chain, the effective detector+readout noise is 2.05$\times 10^{-16}~\mathrm{W/\sqrt{Hz}}$. This noise aliasing will be resolved for the FI using Nyquist inductance in order to reduce the noise bandwidth of the TES (see Piat et al.~\cite{2020.QUBIC.PAPER4}).
\end{itemize}

The QUBIC TD will be shipped to Argentina in mid-2021 and subsequently deployed in its observing site (see section~\ref{Sec:site} for a description). The objective is to conduct a full characterization of bolometric interferometry on the sky throughout 2021. The performance expected from observation with the QUBIC TD is discussed in section~\ref{Sec:TD_perf}.

\subsubsection{The QUBIC Full Instrument\label{FI}}
The upgrade to the FI will happen after one year of operation of the TD and is therefore scheduled for early 2023. It will consist in:
\begin{itemize}
    \item replacing the 64~back-to-back horn array by one with 400 horns which is already manufactured and being characterized at the sub-system level (see Cavaliere~et~al.~\cite{2020.QUBIC.PAPER7} from this series of articles);
    \item replacing the current mirrors by larger mirrors that are already manufactured and characterized (see O'Sullivan et al.~\cite{2020.QUBIC.PAPER8});
    \item upgrading the single 248~TES array by eight of these (four at 150~GHz and four at 220~GHz) achieving two focal planes of 992~TES each. Note that detailed quasi-optical simulations have shown that the 150~GHz optimization of the back-short of the TES is also nearly optimal at 220~GHz allowing us to use identical detector designs for both frequencies~\cite{2020JLTP..200..363P,PerbostPhD}. A new readout electronics will avoid the TD noise aliasing through the addition of Nyquist inductors~\cite{2020.QUBIC.PAPER4}.
    The additional focal plane also requires the installation of a dichroic filter at 45 degrees in between the two focal planes (light-blue in figure~\ref{Fig:qubic_scheme}-left, and in red  in figure~\ref{Fig:qubic_scheme}-right).
\end{itemize} 
The back-to-back horns are common to both frequencies in the QUBIC design. They have been optimized to be single-moded at 150~GHz but are multi-moded in the 220~GHz band (see O'Sullivan et al.~\cite{2020.QUBIC.PAPER8} section~2.3) resulting in a significantly higher throughput at 220~GHz. 

Despite stronger emission from the atmosphere, as can be found in Table~\ref{tab_qubic_td_simulation_parameters}, simulations have shown (see figure~\ref{Fig:bmodes} in  section~\ref{sect_Bmodes}) that this will result in a similar sensitivity to B-modes at 220~GHz and at 150~GHz. 
Several effects contribute to this result. First, considering the intrinsic detector noise (see Piat et al.~\cite{2020.QUBIC.PAPER4}), our noise budget is dominated by the detectors at 150~GHz and has almost equal contributions from photon-noise and detectors at 220~GHz (see table~\ref{tab_qubic_td_simulation_parameters}). Because of this, the net sensitivity loss at 220~GHz due to higher emissivity from the atmosphere is not as large as expected from pure photon-noise. 
We also collect more CMB photons at 220~GHz due to the horns higher throughput. 

As a result, the signal to noise ratio in the TOD at 220~GHz is higher than for a single-moded channel. 
Finally, multimoded horns at 220~GHz also result in a flatter primary beam (see O'Sullivan et al.~\cite{2020.QUBIC.PAPER8} figure~5) than at 150~GHz where the primary beam is Gaussian. This strongly impacts the spatial noise correlation in our maps (see section~\ref{sims}) resulting in a reduction of our power spectra error-bars at the lowest multipoles. This effect occurs for both bands but is stronger at 220~GHz due to the higher amplitude of the synthesized beam secondary peaks.

\subsubsection{The QUBIC site}
\label{Sec:site}
QUBIC will be installed in its final observing site in Argentina at the end of 2021\footnote{It was anticipated to ship the instrument mid-2020 but the global shutdown caused by the COVID19 pandemic induced uncontrolled delays. As a consequence, the date mentioned here is subject to changes depending on the resolution of the COVID19 pandemic crisis.}. The site is located at the Alto Chorillos ($24
^\circ 11'11.7''$ S;
$66^\circ 28'40.8''$ W, altitude of 4869 m a.s.l.) about 45 minutes drive from the city of San Antonio de los Cobres in the Salta Province~\cite{2018BAAA...60..107B}. This site has been studied for mm-wave astronomy for many years as it will also host the LLAMA 12~m antenna (\href{https://www.llamaobservatory.org/en/}{https://www.llamaobservatory.org/en/}), 800~m away from QUBIC. The synergy between QUBIC and LLAMA at the site simplifies  the site preparatory works, logistics and deployment operations.

This site exhibits excellent quality sky for CMB studies: zenith optical depth at 210~GHz $\tau_{210} <0.1$ for 50\% of the time and $<0.2$ for 85\% of the time as well as relatively quiet atmosphere (winds $<6$~ m/s for 50\% of the time). From the LLAMA site-testing data, we have determined an average atmospheric temperature of 270~K with an average emissivity 0.081 and 0.138 at 150 and 220~GHz respectively taken at our average elevation of 50 degrees. These values are assumed for the atmospheric background in the simulations presented in this article\label{atm}. 

\begin{figure}[t]
\centering
\includegraphics[ width = 0.9 \hsize ]{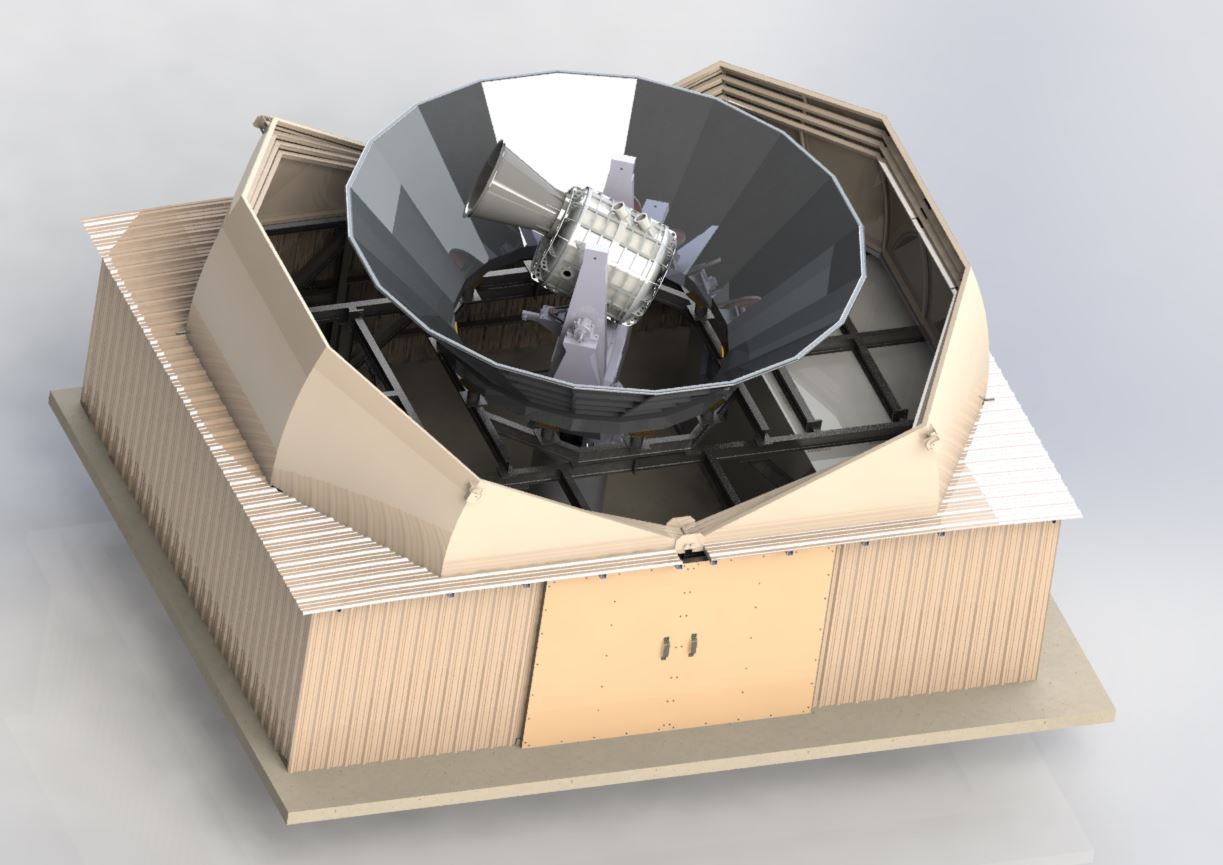}
\caption{Sketch of the instrument, the cryostat on the alt-azimuthal mount as will be installed on the observing site.}
\label{Fig:dome}
\end{figure}

 The whole instrument is oriented to any sky direction with elevation between 30 and 70~degrees (limitations due to Pulse-Tubes-Coolers) and any azimuth by an alt-azimuthal mount on the top of a well-adapted container. A fore-baffle will be placed at the window entrance with an absorptive inner surface in order to increase side lobes rejection for angles larger than 20~degrees from bore-sight direction. Finally, the instrument will be surrounded by a ground-shield in order to minimize brightness contrast between the sky and the ground. Figure~\ref{Fig:dome} shows an artist view of the mounted instrument. The calibration source used for self-calibration (see Torchinsky et al.~\cite{2020.QUBIC.PAPER3} for details) will be installed on a 50~m-high tower (telecommunication-like) placed 50~m North from the instrument. It will directly be seen at normal elevations (without the use of a mirror) in the far field of the instrument and in the direction opposite to the main CMB observations.  

\section{Science Objectives of QUBIC}
\label{sec:sciobj}
QUBIC, being a bolometric interferometer, has distinctive features with respect to traditional imagers designed for observing the CMB and optimized for measuring primordial B-mode polarization. QUBIC is a clean spectral polarimeter well adapted to the purpose of detecting primordial B-modes because of the following main features:
\begin{itemize}
    \item it scans the sky with a synthesized beam formed by the feedhorn array, that exhibits multiple peaks (see Fig.~\ref{Fig:sbqubic} and~\ref{Fig:sb_freq}). Each of the synthesized beam peaks is well approximated by a Gaussian (above -20~dB) with a resolution of 23.5~arcminutes at 150~GHz; 
    \item the angular separation on the sky between the synthesized beam peaks is given by the smallest distance between two peaks and is 8.8~degrees at 150~GHz. This distance scales as a function of frequency within the physical bandwidth of 25\% in such a way that sub-frequency maps can be reconstructed using spectral imaging;
    \item a specific optical design with a polarizing grid before any optical component (except for the Half-Wave-Plate, filters and window) combined with full-power detectors on the focal plane make QUBIC largely immune to cross-polarization.
\end{itemize}
All of these specific features were studied in detail and are incorporated in the forecasts shown in this section. 

\subsection{Data Analysis and Simulations for QUBIC performance forecasts\label{sims}}
\subsubsection{Map-making and noise structure\label{sec_map_making_noise_structure}}
The non-trivial shape of the synthesized beam (figures~\ref{Fig:sbqubic} and~\ref{Fig:sb_freq}) requires a specific map-making method which was developed for QUBIC. It is based on forward modeling to solve the inverse problem of map-making (see Mousset et al.~\cite{2020.QUBIC.PAPER2} for details).  We start from a guess map of the sky and a detailed model of the instrument\footnote{The software 
makes heavy use of the massively parallel libraries developed by P.~Chanial pyoperators~\cite{chanial2012pyoperators} (\href{https://pchanial.github.io/pyoperators/}{https://pchanial.github.io/pyoperators/}) and pysimulators (\href{https://pchanial.github.io/pysimulators/}{https://pchanial.github.io/pysimulators/}).}. The instrument model uses a detailed description of the synthesized beam. At first we can use an idealized instrument model inspired by an ideal synthesized beam as shown in figure~\ref{Fig:sb_freq}, or a more realistic one as simulated including optical aberrations (see figures 11 and 12 in O'Sullivan et al.~\cite{2020.QUBIC.PAPER8}). Then, such a description is expected to be gradually refined during the observation campaigns using information from self-calibration~\cite{2013A&A...550A..59B} in order to incorporate instrumental systematic effects (from optics, electronics) as well observational effects such as ground-pickup or atmospheric contamination. Our software is designed to be able to incorporate a large variety of such systematic effects through the use of a number of time-domain, frequency-domain, or map-domain operators. We observe the guess map at iteration $i$ with the same scanning strategy used with the real data, obtaining simulated TOD that are compared to the real TOD with a $\chi^2$. Using a preconditioned conjugate gradient we modify the guess map in an iterative manner until convergence. The final guess map is the solution of the map-making linear problem. There are two ways of accounting for systematics effects in the map-making: i) directly implementing their action in the operators that produce the simulated TOD in the case we have a model with known parameters from self-calibration and ii) fitting the unknown instrumental systematics parameters (for instance Jones matrices elements~\cite{2007MNRAS.376.1767O}) as extra-parameters, along with the maps pixels, as part of the map-making.

The synthesized beam used for the instrument model during map-making is actually just a set of Dirac functions with the relevant amplitude at the location of the peaks of the synthesized beam (ideal, including optical aberrations or resulting from self-calibration). In such a way, and similarly as with an imager, the map-making does not attempt to deconvolve from the resolution of the peaks, but only from the multiple peaks.

If the synthesized beam model accounts for the realistic frequency dependence, one can use multiple maps at multiple frequencies within the physical bandwidth of the instrument and therefore reconstruct such sub-frequency maps. This is spectral imaging~\cite{2020.QUBIC.PAPER2}. 

Another specificity of this map-making is that the presence of multiple peaks separated by 8.8~degrees on the sky at 150~GHz (6~degrees at 220~GHz) makes QUBIC insensitive to modes on the sky larger than this separation. This occurs because the deconvolution from the multiple peaks relies on the measured signal difference between observations pointing to different directions where the peaks capture different amounts of power. For sky signals at angular scales larger than the angular distance between the peaks, such a difference vanishes. This naturally filters-out large-scale information be it from the sky itself, or from atmospheric gradients. This of course only applies to atmospheric fluctuations at scales larger than the angular distance between peaks in the synthesized beam 8.8 and 6 degrees at 150 and 220 GHz respectively.

To understand the impact of fluctuations on smaller scales we need to run simulations based on dedicated atmosphere measurements taken on site. In appendix~\ref{app_impact_small_scale_atm}, however, we briefly discuss this issue and show how we expect a significant impact only from turbulence cells in a restricted range of scales.

\begin{figure}[!t]
    \centering
    \includegraphics[ width = \hsize ]{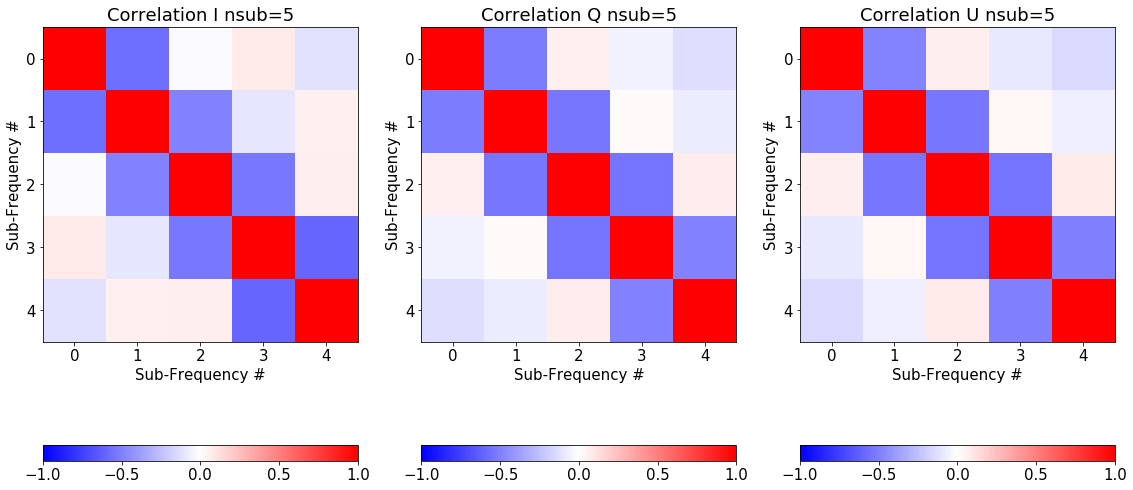}
    \caption{QUBIC sub-bands correlation matrices for I, Q and U Stokes parameters maps obtained from end-to-end simulations at 150 GHz reconstructing the TOD onto 5 sub-bands (equally spaced in log within the physical 150 GHz bandwidth of QUBIC that ranges from 131 to 169 GHz) using spectral imaging and averaged over pixels in the maps. }
    \label{Fig:nunu_corr}
    \end{figure}

A careful study of the noise structure in end-to-end simulated maps shows two significant features that merit further explanation.
    The first is sub-band correlations.  When performing spectral imaging, we show that nearby sub-bands exhibit a significant level of noise anti-correlation~\cite{2020.QUBIC.PAPER2}, as displayed in figure~\ref{Fig:nunu_corr} for the three Stokes parameters at 150~GHz and 5~sub-bands. The anti-correlation is strong with the nearest sub-bands but reduces significantly beyond. Similar correlation matrices are found at 220~GHz.

\begin{figure}[h]
    \centering
    \includegraphics[ width = 0.5
    \hsize ]{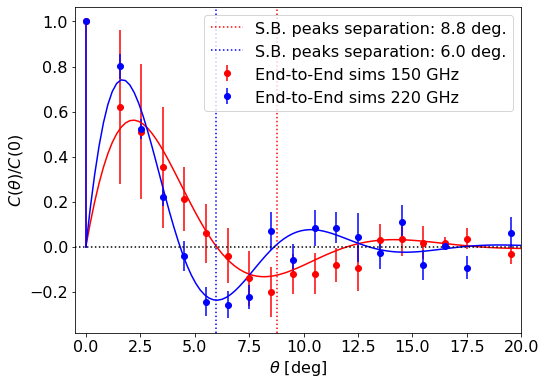}\includegraphics[ width = 0.5 \hsize]{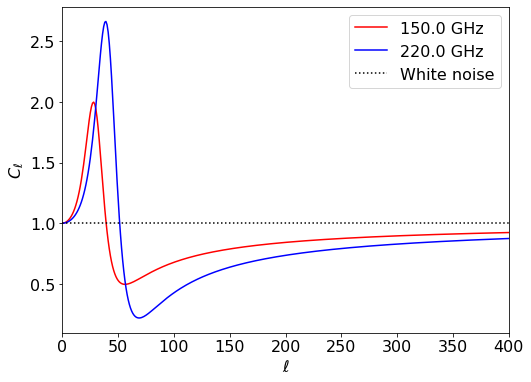}
    \caption{{\bf (left)} QUBIC spatial noise 2pt-correlation function obtained from end-to-end simulations normalized by the variance in the maps $C(\theta=0)$. The solid lines show an adjustment by a sine-wave modulated by an exponential with a Dirac function at $\theta=0$ (the noise variance in the maps). The maximum anti-correlation is found as expected at the scale of the angular distance between two peaks of the synthesized beam (S.B.). The amplitude of the correlation is higher at 220~GHz than at 150~GHz because of the top-hat shape of the primary beam at 220~GHz. {\bf (right)} Spatial noise correlation converted to multipole space. The straight-line at $C_\ell=1$ shows the expected shape for white noise. The noise correlation results in a reduction of the noise for multipoles larger than $\sim 40-50$, that is for angular scales $\lessapprox\theta_\mathrm{peaks}$ (angular separation between the synthesized beam peaks). At lower multipoles (larger angular scales), we observe an increase of the noise. This is an advantage for measuring the recombination peak around $\ell=100$ as discussed in section~\ref{sect_Bmodes}}.
    \label{Fig:space_corr}
    \end{figure}

    The second feature worth noticing is spatial correlations.  Map-making with a multiply-peaked synthesized beam involves partial deconvolution because a given time sample in a detector's TOD receives power from distinct pixels in the sky with weights given by the shape of the synthesized beam.  As a result, we expect significant spatial noise correlations in our maps. This is confirmed by end-to-end simulations as shown in the left panel of figure~\ref{Fig:space_corr}. Anti-correlation peaks, are expected, at an angle corresponding to the angular separation between the peaks in the synthesized beam ($\theta_{\mathrm{peaks}}$=8.8~degrees at 150~GHz and 6 degrees at 220~GHz). A similar 2pt-correlation function is found at 220~GHz, but with even higher correlation amplitude because the secondary peaks are higher due to the top-hat shape of the primary beam resulting from multimode optics at 220~GHz (see figure~4 in O'Sullivan et al.~\cite{2020.QUBIC.PAPER8}).

    In the right panel of figure~\ref{Fig:space_corr} we display  the spherical harmonics transform of the 2pt-correlation function:
    \begin{equation}
        C_\ell = 2\pi\int_{-1}^{1}C(x)P_\ell(x)\mathrm{d}x
    \end{equation}
    where $x=\cos\theta$ and $P_\ell$ are the Legendre polynomials.
    This is our noise angular power spectrum which corresponds to the equivalent for QUBIC of typical white noise for a classical imager. The shape of this noise in Harmonic Space exhibits an excess with respect to white noise at very large scales (small multipoles, below $\ell=40$  at 150 GHz and $\ell=50$ at 220 GHz) and a significant reduction at smaller angular scales (larger multipoles). The scale of this transition is determined by the angular distance between peaks in the synthesized beam\footnote{It is however not strictly equal to $\pi/\theta_\mathrm{peaks}$ because of the shape of the 2pt-correlation function and the non-trivial correspondence between angles and multipoles.}. Angular scales $\gtrapprox\theta_\mathrm{peaks}$ are not well constrained due to the presence of the multiple peaks that are effectively deconvolved during the map-making. Conversely, angular scales smaller than this angular separation see their noise significantly reduced thanks to the positive correlation of the noise at these angles. 
    Because these angular scales correspond to those of the recombination peak in the B-mode spectrum, this specific noise feature for Bolometric Interferometry turns out to be a significant advantage for detecting primordial B-modes. This is discussed with more details from Monte-Carlo simulations in section~\ref{sect_Bmodes} and visible in Figure~\ref{Fig:bmodes}. Also, because our noise is not white, the RMS in the maps does not have direct significance. In our case, it is more meaningful to measure the noise level in the angular power-spectrum as we will detail in section~\ref{sect_Bmodes}.
    
\subsubsection{The QUBIC Fast Simulator}
The peculiar noise structure of the QUBIC maps has been studied in detail using a number\footnote{40 end-to-end simulations were used for each of the considered configurations.} of end-to-end simulations run on supercomputers as they have large memory requirements\footnote{With a 156.25~Hz sampling rate, QUBIC produces $156.25 \times 3600 \times 24 \times 1024 \times 2 \times 8 /1e9 \simeq 220$~GB/day.}. We extract from these simulations the main features of the noise discussed above:
\begin{itemize}
    \item noise scaling as a function of normalized coverage;
    \item correlations between reconstructed sub-bands (see figure~\ref{Fig:nunu_corr});
    \item spatial (pixel-pixel) noise correlation measured on maps corrected for the noise scaling with respect to coverage;
\end{itemize}
We have built a "Fast Simulator" that directly produces maps with theses features: 
\begin{enumerate}
    \item we start by creating (in harmonic space) noise maps with the observed spatial correlation (see figure~\ref{Fig:space_corr}),
    \item we then make linear combinations of these noise maps in order to have I, Q and U maps for each sub-band with the appropriate correlation matrix (see figure~\ref{Fig:nunu_corr}),
    \item finally, we scale the noise in the maps according to the scaling with respect to coverage. 
\end{enumerate} 
The overall noise normalization is adjusted to match that of the end-to-end simulations with the same integration time. We have checked in detail the accuracy of the Fast Simulator by performing the same noise structure analysis on the output maps and verifying that they lead to the same noise modeling as with the end-to-end simulations.

The Fast Simulator allows for fast production of maps with large-number statistics (thousands of realizations) and has been used extensively for the forecasts presented in this article.

We have used a simplified sky coverage obtained using random pointings on the sky from one time sample to another, reproducing the same sky fraction as the anticipated sky coverage shown in figure~\ref{Fig:coverage}. While this allows obtaining a fast and efficient coverage of the QUBIC observed sky, it prevents one from simulating actual $1/f$ noise from atmospheric or any time-domain instrumental fluctuations as successive time samples do not correspond to nearby pointings as in a more realistic scanning strategy. As a consequence,
the simulations presented in this article implicitly assume a stable atmosphere with no $1/f$ noise, but account for the average loading from the atmosphere (see section~\ref{overview} for a discussion of the seasonal and diurnal variations). 
We will include realistic atmospheric fluctuations in our simulations when more detailed information than the average emissivity and temperature will be available from the TD data on the sky. In addition, as discussed in the second paragraph of this subsection, the shape of the QUBIC synthesized beam implies a low sensitivity on scales much larger than the angular distance between our multiple peaks, which will significantly reduce the impact of the (mostly large scales) atmospheric fluctuations.

No instrumental systematics are considered in this article, they will be included in future studies as well as how they can be mitigated through self-calibration. All results presented in this article should therefore be considered as idealized and intended to estimate the ultimate sensitivity achievable by a perfect instrument observing a stable atmosphere.

\subsection{The quest for CMB B-modes}
\label{sec:bmodeoverview}
  
\subsubsection{Overview}\label{overview}
We have performed simulations for a three-year observation of the sky using the QUBIC FI on a sky without any foregrounds, but with realistic instrumental noise~\cite{2020.QUBIC.PAPER3,2020.QUBIC.PAPER4}) and with atmospheric background noise (assumed to be stable). The latter has been obtained from measurements performed over 3 years at the QUBIC site in Argentina from a tipper at 210~GHz used for the LLAMA radiotelescope\footnote{ \href{https://www.llamaobservatory.org/}{https://www.llamaobservatory.org/}} to be installed near QUBIC. Atmospheric background is averaged over the 9~best months of the year and corresponds to an atmospheric temperature of 270~K and emissivities 0.081 and 0.138 at 150 and 220~GHz respectively for an average observation elevation of 50 degrees. 

Changing the atmospheric parameters according to maximal diurnal and seasonal variations induces 15 and 25\% change in the photon noise at 150 and 220 GHz respectively (with respect to the numbers given in Table~\ref{tab_qubic_td_simulation_parameters}). This corresponds to a negligible change of sensitivity for the TD because our noise is dominated by that of the detectors. For the FI, the change in the total noise is 5\% at 150 GHz and 20\% at  220 GHz. As said before, more detailed simulations including the measured atmospheric fluctuations will be performed when these measurement are available with the instrument.

We have used the ``Fast Simulator'' described above to produce thousands of realizations of the noise in the maps incorporating the peculiar noise structure (spatial variations of the noise as a function of coverage as well as spatial and sub-band correlations). The parameters of our simulation are summarized in table~\ref{tab_qubic_td_simulation_parameters}.

\begin{table}[t]
    \renewcommand{\arraystretch}{1.}
    \begin{center}
        \caption{\label{tab_qubic_td_simulation_parameters}Main instrumental and simulation parameters used in our computations. The noise value for TD is measured from the TES calibration data~\cite{2020.QUBIC.PAPER3} while that for the FI is the intrinsic TES noise and assumes reduction of the noise aliasing in the readout chain found in the TD. We have used average atmospheric parameters at 50 degrees elevation accounting for maximal seasonal and diurnal variations corresponds to negligible change for the total noise for the TD, 5\% and 20\% for the FI at 150 and 220 GHz respectively (see text).}
        \begin{tabular}{p{6cm} p{3.7cm} m{3.8cm}}
            \hline
            Parameter& Value TD & Value FI \\
            \hline
            \hline
            Detector noise [$\mathrm{W/\sqrt{Hz}}$]\dotfill & 2.05$\times 10^{-16}$ & 4.7$\times 10^{-17}$ \\
            
            \hline
            Atmosphere$^1$ temperature [K]\dotfill& 270 &270\\
            Atmosphere emissivity$^1$ at 150\,GHz\dotfill& 0.081 &0.081\\
            Photon noise [$\mathrm{W/\sqrt{Hz}}$]\dotfill & 2.6$\times 10^{-17}$~(150 GHz) & 3.1$\times 10^{-17}$~(150 GHz),\hfill\break 1.17$\times 10^{-16}$(220 GHz)\\
            \hline
            Total noise [$\mathrm{W/\sqrt{Hz}}$]\dotfill & 2.06$\times 10^{-16}$(150 GHz) & 5.7$\times 10^{-17}$~(150 GHz),\hfill\break 1.26$\times 10^{-16}$~(220 GHz)\\
            \hline
            Cumulated observation time [years]\dotfill & 1 & 3  \\
            \hline
            $r$ upper-limit (68\% C.L., No FG) &- & 0.021~(150~GHz),\hfill\break 0.023~(220~GHz),\hfill\break 0.015~(Combined)\\
            \hline
            \multicolumn{3}{l}{$^1$The atmosphere is considered perfectly stable.}\\         \end{tabular}
    \end{center}
\end{table}

For each realization, we have used NaMaster~\footnote{\href{https://github.com/LSSTDESC/NaMaster}{https://github.com/LSSTDESC/NaMaster}}~\cite{2019MNRAS.484.4127A} to compute pure TT, EE, TE and BB power spectra on the residual maps (therefore noise-only maps) in order to compute the expected noise on the power spectra. Exploring various values for the minimum multipole $\ell_{min}$, the size of the multipole bins $\Delta_\ell$ and minimum value of the relative coverage (normalized to 1 at maximum) of the sky that defines the region of the sky we keep for analysis, $\mathrm{Cov}_c$, we have found the best configuration to be $\ell_\mathrm{min}=40$, $\Delta_\ell=30$ and $\mathrm{Cov}_c=0.1$ (keeping all pixels with relative coverage above 0.1) at 150 GHz. We have kept the same configuration for the 220~GHz for the sake of simplicity. 

\subsubsection{QUBIC Full Instrument expected performance}
\label{sect_Bmodes}
\begin{figure}[!t]
\centering
\includegraphics[ width = 0.49 \hsize ]{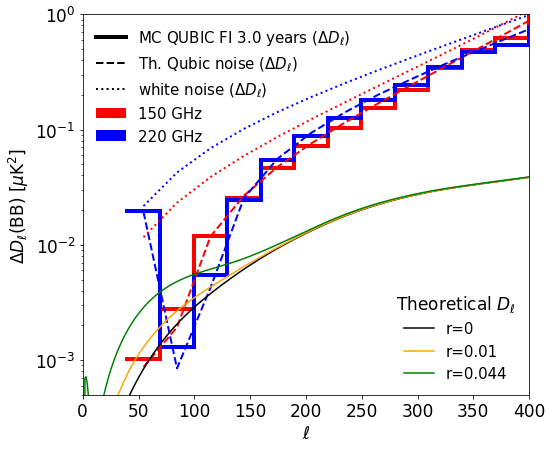}\includegraphics[ width = 0.49\hsize ]{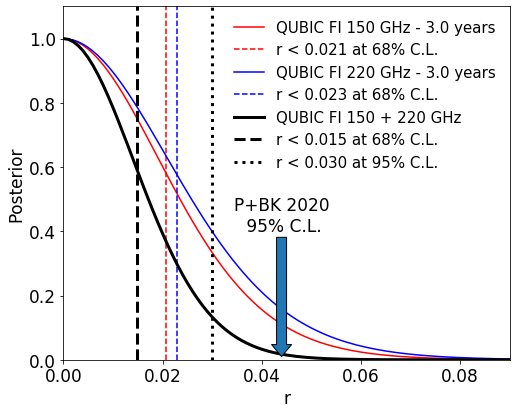}
\caption{{\bf (left)} BB power spectrum error-bars $\Delta D_\ell$ on $D_\ell = \frac{\ell(\ell+1)}{2\pi}C_\ell$ on residual maps from the Fast-Simulator Monte-Carlo in the absence of foregrounds, atmospheric fluctuations and instrument systematic effects for an integration time on the sky of three years from the site in San Antonio de los Cobres, Argentina. As these are calculated on residual maps, they do not incorporate sample variance and only refer to the instrumental noise.
The reduction at low-$\ell$ of our error-bars with respect to theoretical white-noise (dotted lines) is clearly visible and in agreement with the expected shape from the 2pt correlation function in figure~\ref{Fig:space_corr} (dashed lines). The difference in the first bin is discussed in the text.
{\bf (right)} Posterior likelihood on the $r$ (assuming no foregrounds) using QUBIC FI (two bands) with three years integration on the sky (including both noise and sample variance). The latest Planck+Bicep/Keck constraint from Tristram et al.~\cite{Tristram_r} is shown with the blue arrow}
\label{Fig:bmodes}
\end{figure}

As remarked before, measuring the RMS of a map in the case of non-white noise is meaningless, and the actual measurement of the effective depth of our maps is done in $\ell$-space. The resulting uncertainties from the Monte-Carlo on the BB polarization power spectrum ($\Delta D_\ell$) are shown in the left panel of figure~\ref{Fig:bmodes}. 
We have also displayed theoretical B-mode power spectra $D_\ell$ (including lensing) for $r$=0, 0.01, and 0.044 (current best upper limit from Tristram et al.~\cite{Tristram_r}). Besides the QUBIC error bars, we also plot the expected shape for white noise (dotted lines) and for the QUBIC noise (dashed lines) from figure~\ref{Fig:space_corr}. 
In both cases the theoretical error-bars are obtained through the well known formula:
\begin{equation}
    \Delta D_\ell^\mathrm{th} = \frac{\ell(\ell+1)}{2\pi}\times \sqrt{\frac{2}{(2\ell+1)f_\mathrm{sky}\Delta\ell}}\times \frac{1}{B_\ell^2}\times \frac{1}{W_\ell^2} \times C_\ell^\mathrm{noise}
\end{equation}
where $f_\mathrm{sky}$ is the fraction of the sky used for analysis ($f_\mathrm{sky}=0.015$ in our case), $\Delta\ell$ is the width of the $\ell$-space binning ($\Delta\ell=30$ in our case). $B_\ell$ is the beam transfer function and $W_\ell$ is that of the Healpix pixellisation ($N_\mathrm{side}=256$). $C_\ell^\mathrm{noise}$ is the expected shape for the noise power spectrum that can be either a constant in the case of white noise or a constant multiplying the shape from figure~\ref{Fig:space_corr} (right) in the case of QUBIC. We performed a fit of the above noise normalization for QUBIC to our Monte-Carlo error-bars leading to 2.7~and 3.7~$\mu\mathrm{K.arcmin}$ at 150~and 220~GHz respectively. This is shown as dashed lines in figure~\ref{Fig:bmodes}. However these numbers are hardly comparable with the case of a standard imager for which the noise is white: for each frequency we overplot the expected shape for white noise with the same normalization. The significant noise reduction with respect to the white noise case is particularly visible at the scales of the recombination peak near $\ell=100$, giving QUBIC an enhanced sensitivity at those scales. At scales larger than the separation between peaks in the synthesized beam however, the error-bars increase sharply.
The first bin at 220~GHz exhibits significantly larger error-bars for all spectra. This is not surprising as we have kept the same $\ell_{min}=40$ for both channels. In reality, the multiple-peaked shape of our synthesized beam is such that we have little sensitivity to multipoles corresponding to angular scales larger than the distance between the peaks (8.8~and 6~degrees at 150~and 220~GHz respectively). As a result, the optimal $\ell_{min}$ at 150~GHz is slightly too low for 220~GHz, resulting in larger error-bars for the first bin. This will be optimized when analyzing real data.

The right panel of figure~\ref{Fig:bmodes} shows the posterior on the tensor-to-scalar ratio which, in the absence of foregrounds, was the only free-parameter for this power-spectrum-based likelihood (simple $\chi^2$ accounting for sample variance~\cite{hamimeche-lewis}) with all parameters but $r$ fixed to their fiducial values\footnote{We have used a fiducial cosmology with parameters [$h=0.675$, $\Omega_b h^2=0.022$, $\Omega_c h^2=0.122$, $\Omega_k=0$, $\tau=0.06$, $A_s=2e-9$, $n_s=0.965$]}.
We calculate the likelihood at 150 and 220 GHz separately as well as jointly. These simulations show that QUBIC  has the statistical power (without foregrounds, atmospheric fluctuations and systematics) to constrain the B-modes down to a tensor-to-scalar ratio $r<0.015$ at 68\% C.L. ($r<0.03$ at 95\% C.L.) with three years integration on the sky from our site in Argentina. 

In the presence of foregrounds,  the numbers above are to be understood as our statistical sensitivity to effective B-modes including the contribution from primordial tensors as well as dust polarization. 
Component separation has not been included in the current forecasts and will be investigated in details in a future publication. In our study we have focused only on dust contamination, neglecting a possible contribution from synchrotron. This hypothesis is supported by two facts: (i) QUBIC will search for cosmological B-modes in a well defined sky patch that will correspond roughly to the one shown in figure~\ref{Fig:coverage}, and (ii) in this sky patch the systematic error on $r$ from synchrotron contamination at 150\,GHz is below 0.005 (see figure~6 in Krachmalnicoff et al.~\cite{krachmalnicoff2016}), so well below our target $r<0.021$. Moreover, QUBIC will have the capability to check for any residual foregrounds contamination in the obtained tensor-to-scalar ratio by exploiting its spectral imaging features as described in section~\ref{sss:dust_residuals}. Finally, in the QUBIC data analysis we will also exploit the wealth of data available at low frequency (WMAP, Planck, C-Bass, QUIJOTE and other data that will be publicly available) to further improve the robustness of our final results.

\subsection{The foregrounds challenge}
\label{sect:foreground_challenge}

    \subsubsection{Overview}
\label{sec_fg_challenge_overview}

    Instrumental systematic effects and polarized astrophysical foregrounds are the main challenges to current and future generation \bmode\ experiments. Indeed, polarized emissions from foregrounds are brighter than the \bmode\ signal over the entire sky, and the only way to separate the CMB signal from the foregrounds is to make measurements at several frequencies in a range from a few GHz to several hundreds GHz.

The polarized foregrounds are dominated by the synchrotron emission, generated by cosmic ray electrons spiraling around the Galactic magnetic field, and by Galactic dust emission, caused by magnetized grains, heated by starlight and aligned with the direction of the Galactic magnetic field (\cite{planck_iv_2018} and references therein). The frequency scaling law of the synchrotron emission reflects the energy distribution of the electrons and can be described with a power law characterized by a spectral index $\beta_\mathrm{synch}\sim -3$. The dust behaves like a gray-body with a temperature of $\sim$18~K and an emissivity with spectral index $\beta_\mathrm{dust}\sim 1.5$~\cite{planck-2015-xxvi, planck_iv_2018}. These parameters, however, vary across the sky and display spatial correlations that are essentially unknown.

The scenario is made even more complex also by other emissions that could impact \bmode\ measurements. The anomalous microwave emission (AME), for example, is correlated with the dust and emits at low frequencies as a result of spinning grains~\cite{genova-santos-2017}.  Carbon-monoxide (CO) lines~\cite{puglisi2017} correlate with Galactic gas clouds.  Extra-Galactic foregrounds are generated by radio and infrared sources in a wide frequency range between one and several hundreds of GHz, with brightness temperatures that may decrease (radio sources) or increase (infra-red sources) with frequency~\cite{planck-2015-xxvi, puglisi2018}. 



A thorough analysis of WMAP, Planck and S-PASS data carried out by Krachmalnicoff~et~al.~\cite{krachmalnicoff2016, krachmalnicoff2018} has shown that there is no sky region that is clean enough from foregrounds contamination to allow a significant \bmodes\ detection below $r = 0.01$. Furthermore they found a correlation of the order of 10\% between synchrotron and dust emissions on large angular scales. This clearly shows that the control of foregrounds and of its detailed spectral behavior is mandatory for any \bmode\ experiment.

Another compelling piece of evidence of the impact of foregrounds is provided by figure~34 of Planck Collaboration~\cite{planck_iv_2018}, which shows that if $r<0.05$ then there is essentially no region in $\ell$ space and/or in frequency where the cosmological \bmodes\ are brighter than the foregrounds, so that the final uncertainty is dominated by the residuals after component separation.

If, on the one hand, the issue of space coverage and resolution will be addressed by stage-3~\cite{Polarbear2, AdvACT2016, Polarbear+simons2016, 2019JCAP...02..056A} and stage-4 CMB experiments~\cite{cmbs4}, on the other hand frequency resolution requires a different approach in the instrument design. 
Thanks to the ability to discriminate sub-bands in each frequency band, QUBIC offers an alternative to other approaches to foreground mitigation that will provide complementary information to the traditional widely separated observation bands.

\begin{figure}[t]
    \begin{center}
        \includegraphics[width=15cm]{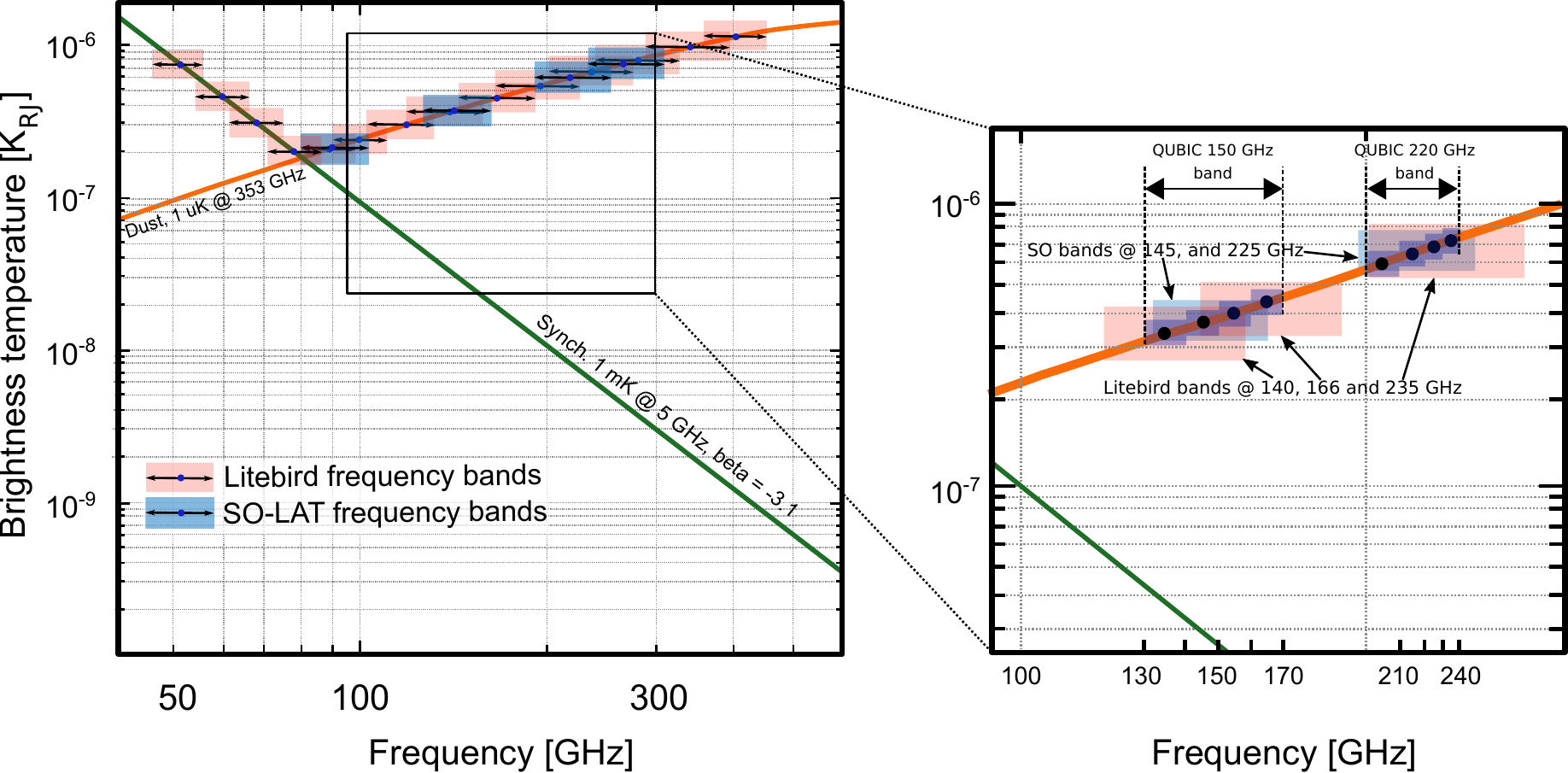}
          \caption{\label{fig_frequency_resolution}Improvement in frequency resolution achievable with the QUBIC spectral imaging capabilities. (\textbf{left}): polarized dust emission and synchrotron in brightness temperature units. Simons Observatory~\cite{2019JCAP...02..056A} and LiteBIRD~\cite{suzuki2018litebird, hazumi2019} frequency bands are overplotted on the synchrotron and dust spectra. (\textbf{right}): expansion of the region around QUBIC frequencies. We can appreciate the $\sim$1~GHz frequency resolution provided by spectral imaging (here with four sub-bands in each wide-band).}
    \end{center}
\end{figure}

Figure~\ref{fig_frequency_resolution} highlights the potential of QUBIC in providing enhanced frequency resolution compared to state-of-the-art and future experiments. In the left panel we see a spectrum of the polarized dust and synchrotron emissions in brightness temperature units. Overplotted to the dust line we see the frequency bands of the Simons Observatory~\cite{2019JCAP...02..056A}, while in the middle panel we see the bands expected for the LiteBIRD satellite~\cite{hazumi2019}. The right panel is an expansion of the region around the two QUBIC bands centered at 150 and 220~GHz, where we appreciate the improvement in frequency resolution achievable with QUBIC (here with four sub-bands in each of our bands).

A relevant feature of the spectro-imaging power of QUBIC is the ability to recognize the presence of foreground residuals remaining in the data after component separation. This is discussed in detail in section~\ref{sss:dust_residuals}.

\subsubsection{QUBIC Technological Demonstrator expected performance}
\label{Sec:TD_perf}

    In this section we discuss the expected scientific performance of the QUBIC TD. Our prototype will prove its spectral imaging power by observing the brightest regions of our Galaxy and reconstructing the spectrum of the emission in total intensity and polarization with a frequency resolution of $\sim$4\,GHz. The full demonstration on a low-foreground CMB field will require higher sensitivity and will be done with the FI.

In our assessment we have simulated the observation of a circular 15 degrees-wide sky patch centered at $(\ell,b)=(0,0)$ and containing various combinations of foregrounds: (i) interstellar dust emission, simulated with the PySM\footnote{See \url{https://pysm-public.readthedocs.io/en/latest/models.html} for details about the used models} \texttt{d1} model, synchrotron emission, simulated with the PySM \texttt{s1} model, and (iii) the sum of the two.

\begin{figure}[t]
    \begin{center}
        \includegraphics[width=0.9\hsize]{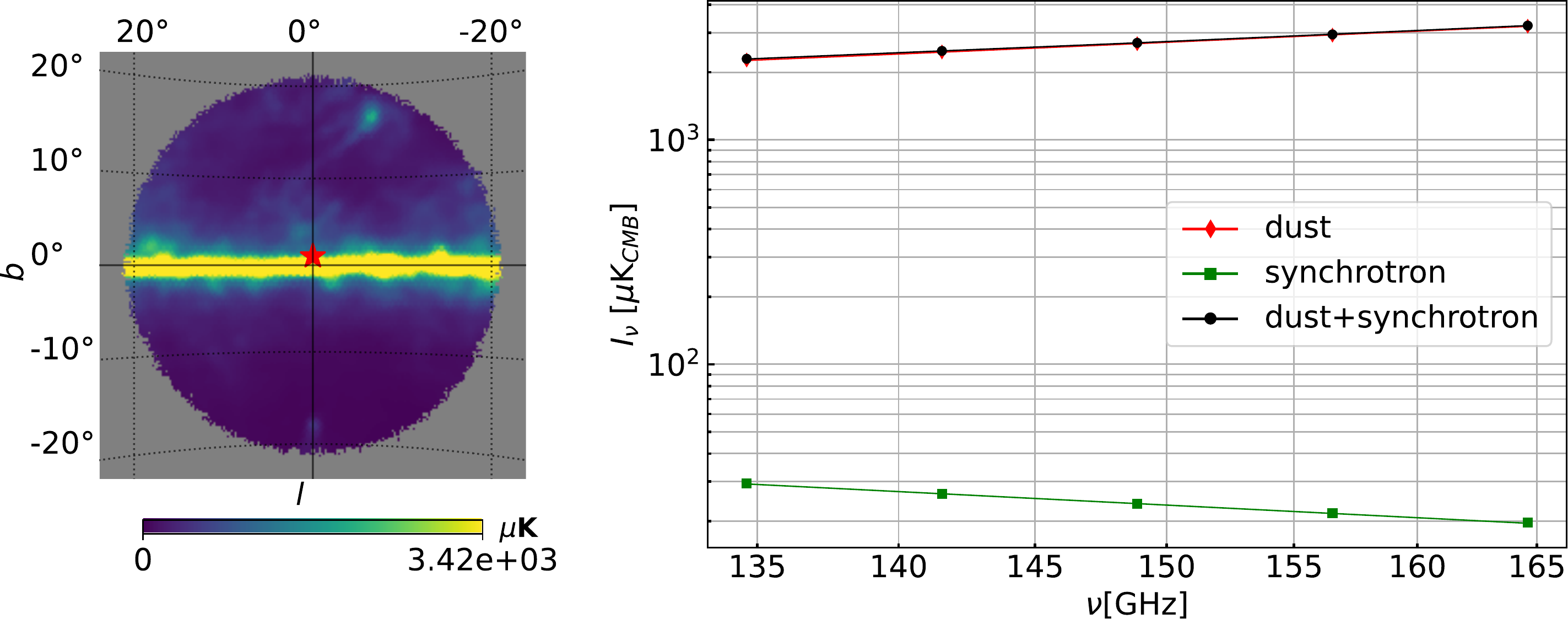}
        \caption{\label{fig_true_sky}(\textbf{left}): HEALPix intensity map of the input sky convolved with the instrument beam ($N_\mathrm{side}=256$). The red star represents a pixel for which we display the SED of emission in the right panel. (\textbf{right}): SED in five QUBIC sub-bands of the input model sky emission computed in the pixel shown in the left panel. Here we consider three cases: dust emission only (red dots), synchrotron emission only (green dots ranging from $\sim 30~\mu$K at 135 GHz to $\sim20~\mu$K at 165 GHz) and sum of the two (black dots).}
    \end{center}
\end{figure}

We have simulated the input sky considering 15 equally-spaced frequencies in the 25\% bandwidth around the 150\,GHz central frequency. In the left panel of figure~\ref{fig_true_sky} we show a HEALPix map ($N_\mathrm{side}=256$) of the input sky convolved at the instrument angular resolution (width of the peaks in the synthesized beam), while in the right panel we plot the spectral energy distribution (SED) for the three considered cases in the pixel marked in red on the map. The dots correspond to the five sub-frequencies reconstructed by QUBIC. From the figure we can already appreciate the fact that in the QUBIC frequency band the synchrotron amplitude is more than two decades below the dust providing a very small contribution to the overall foreground emission.

\begin{figure}[!p]
        \begin{center}
            \includegraphics[width=\hsize]{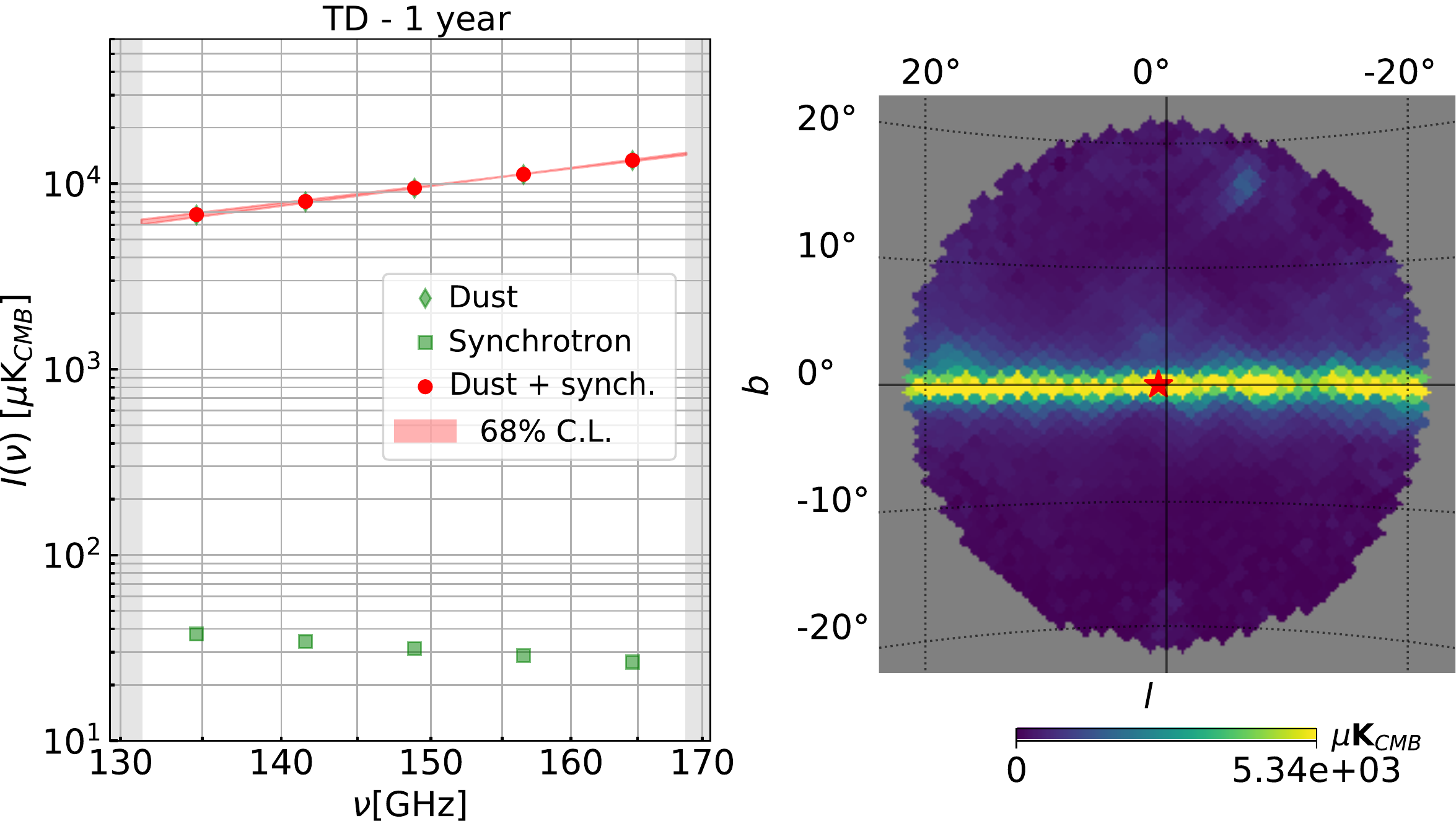}
            \mbox{}\\
            \includegraphics[width=\hsize]{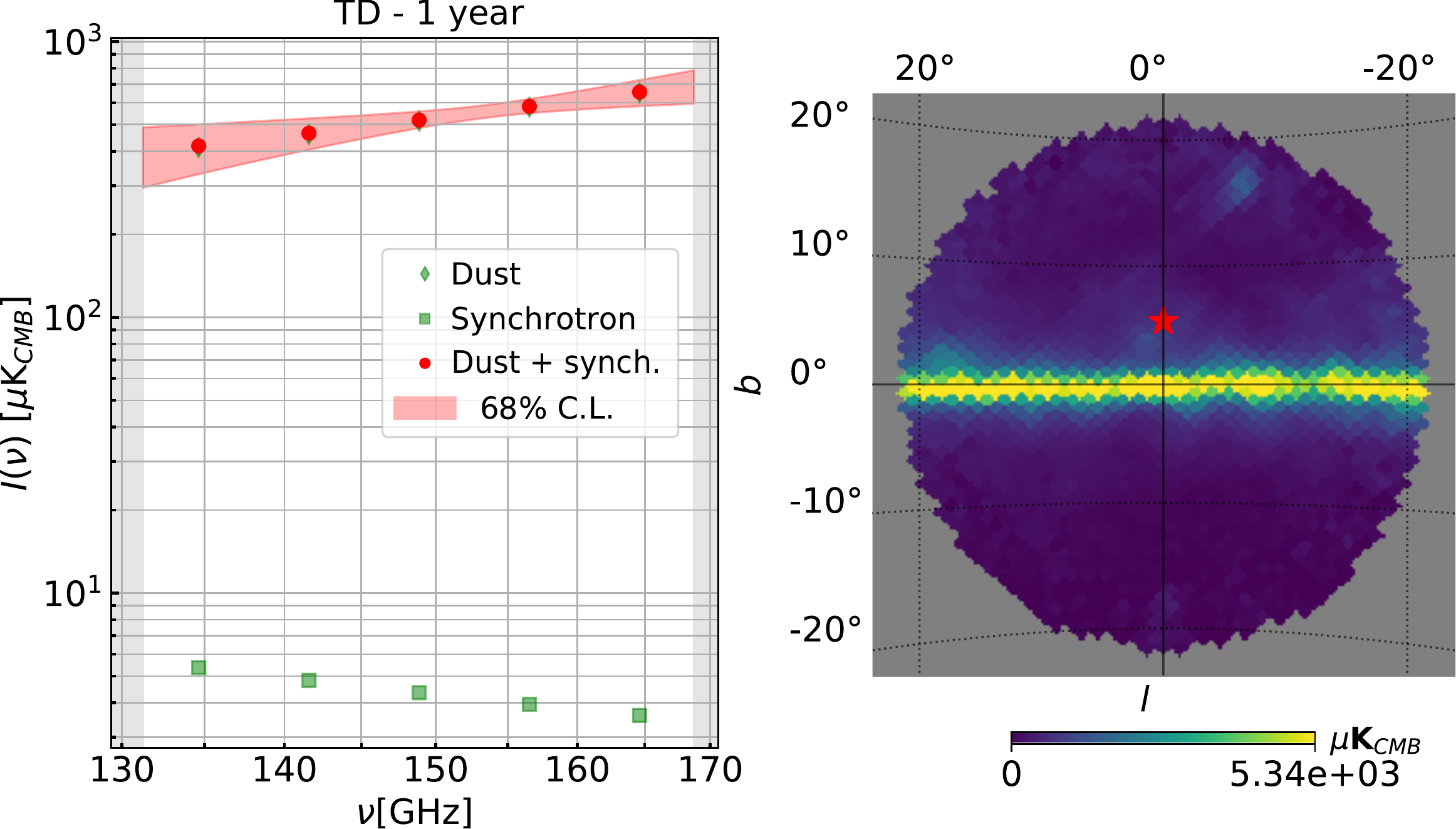}
            \caption{\label{fig_sed_conv_vs_reconstruct_I}Reconstruction of the interstellar dust SED in total intensity with the QUBIC TD in the 150\,GHz band. The two rows show the result in $\sim 1$\,degree pixels ($N_\mathrm{side} = 64$) at two different distances from the Galactic center. In the left column, we show the input values for the selected pixel (red circles) as well as the 68\% C.L. region for the measured SED (see text). The right column shows the reconstructed map. The red marks indicate the pixels chosen for the SED computation. The maps are in Galactic coordinates, centered towards (0,0) with grid spacing in latitude and longitude 10 and 20 degrees respectively.}
        \end{center}
    \end{figure}

\begin{figure}[!p]
        \begin{center}
            \includegraphics[width=\hsize]{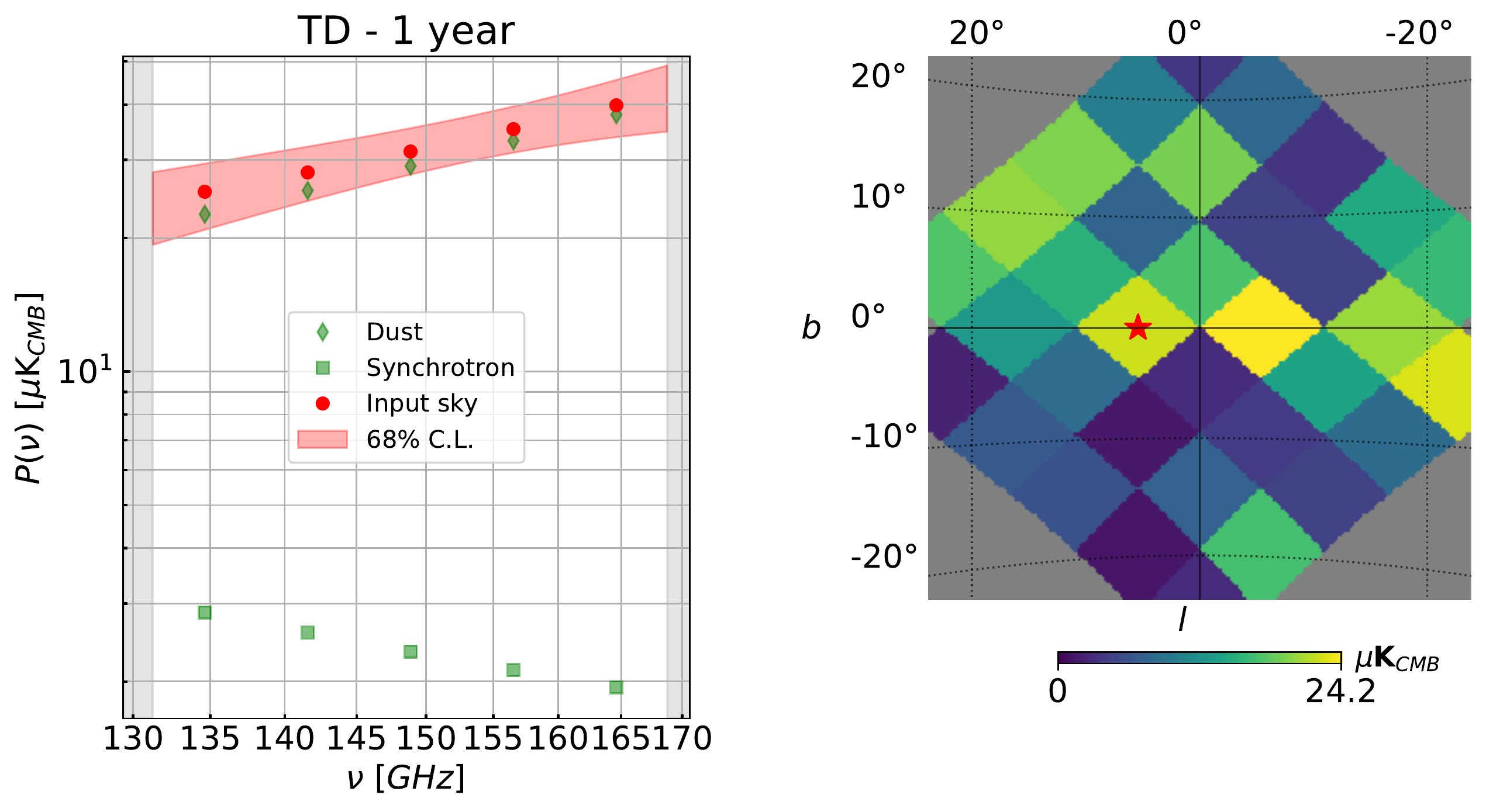}
            \mbox{}\\
            \includegraphics[width=\hsize]{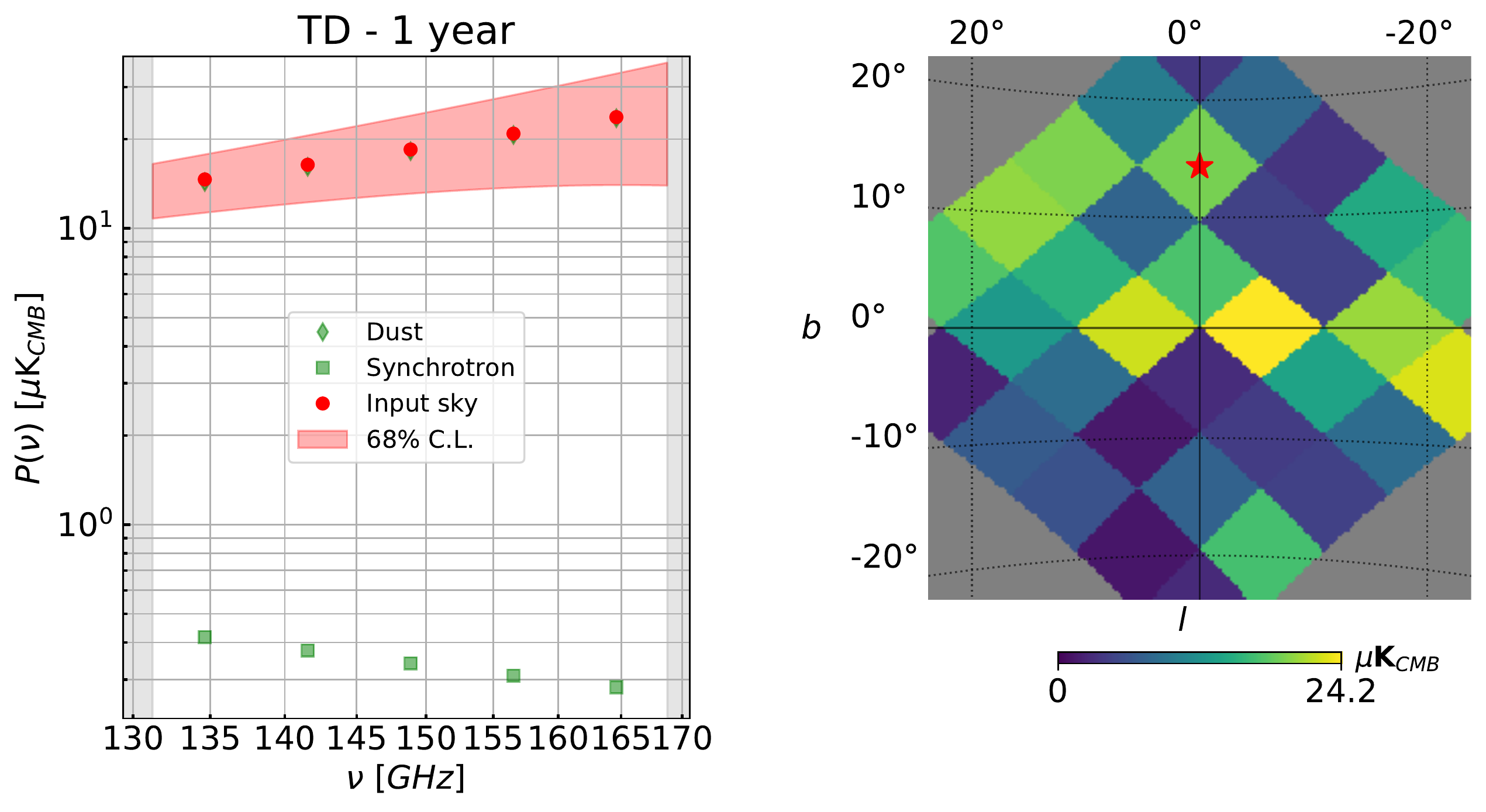}
            \caption{\label{fig_sed_conv_vs_reconstruct_P}Reconstruction of the interstellar dust SED in polarization with the QUBIC TD in the 150 GHz band. The two rows show the result in $\sim 7.3$\,degree pixels ($N_\mathrm{side} = 8$) at two different distances from the Galactic center. In the left column, we show the input values for the selected pixel (red circles) as well as the 68\% C.L. region for the measured SED (see text). The right column shows the reconstructed map. The red marks indicate the pixels chosen for the SED computation. The maps are in Galactic coordinates, centered towards (0,0) with grid spacing in latitude and longitude 10 and 20 degrees respectively.}
        \end{center}
    \end{figure}

\paragraph{Simulation procedure.} First we have calculated 100 realizations of the reconstructed maps of the observed sky patch using the fast simulation pipeline described in section~\ref{sims} with $N_\mathrm{side} = 256$ and degrading the pixel resolution to $N_\mathrm{side} = 64$ for total intensity maps and $N_\mathrm{side}=8$ for polarization maps. Then, for each pixel we have computed the average $I(\nu_j)$, with its 68\% confidence interval for each sub-band frequency $\nu_j$. We calculated the confidence interval by modelling the SED as a modified black-body: 
\begin{equation}\label{modifiedBB}
    I(\nu) = a\times B_\nu(T_\mathrm{dust}=19.6\,\mathrm{K})\times\left(\frac{\nu}{353\,\mathrm{GHz}}\right)^\beta,
\end{equation}
then we performed a Monte-Carlo-Markov-Chain exploration of the posterior likelihood of this model given our full band-band covariance matrix (see figure~\ref{Fig:nunu_corr}). This provides us with a chain of $(a,\beta)$ parameters that sample the likelihood. From these samples, we then calculate the 68\% confidence interval for the model at each frequency in the band, obtaining the light red areas represented in figures~\ref{fig_sed_conv_vs_reconstruct_I} and \ref{fig_sed_conv_vs_reconstruct_P}.

Here we want to underline that in the model used to fit the data we did not consider the synchrotron in addition to the dust emission because, as shown in figure~\ref{fig_true_sky}, the synchrotron emission can be considered negligible in the QUBIC TD frequency band. This is further confirmed in the results displayed in figures~\ref{fig_sed_conv_vs_reconstruct_I} through \ref{fig:sed-fi} where we see that the input sky SED falls into recovered confidence interval independently of the presence of the synchrotron emission.

We verified that the shape of the confidence interval does not depend on the model we assumed for the MCMC exploration. Indeed, we obtain almost identical regions with a second order polynomial instead of the modified power law\footnote{It is worth noting that the angular resolution of our reconstructed maps in each sub-band is not constant and improves with frequency~\cite{2020.QUBIC.PAPER2}. As a result, fitting a modified power law without accounting for this change of resolution does not lead to a $\beta$ parameter that can be compared with the usual $\beta$ dust spectral index that needs to be corrected for varying angular resolution~\cite{irfan2019determining}. This is not a problem here as this analysis is just intended to show our ability to measure SED with spectro-imaging.}. In table~\ref{tab_qubic_td_simulation_parameters} we list the main parameters used in our simulations.

\paragraph{Results.}
    In figures~\ref{fig_sed_conv_vs_reconstruct_I} and \ref{fig_sed_conv_vs_reconstruct_P} we show how we can reconstruct the SED of the dust emission in total intensity and polarization with one year of sky integration with the QUBIC TD in its 150\,GHz band. More in detail, in the left column we show the dust SED reconstructed in particular pixels with its 68\% confidence interval plotted together with the input sky SED for the dust, synchrotron and dust$+$synchrotron cases. In the  right columns we show the reconstructed maps with the corresponding pixel highlighted with a red mark.
     
    Our results show that with one year sky integration with the QUBIC TD it is possible to reconstruct the SED of the dust emission on angular scales of the order of $\sim$1\,degree and frequency resolution of $\sim$4\,GHz in total intensity. We can also detect the dust emission SED in polarization on larger scales ($\sim$7\,degrees) close to the Galactic plane, where the signal intensity is larger.

\subsubsection{QUBIC Full Instrument expected performance}
\begin{figure}[t]
    \begin{center}
    \includegraphics[width=\hsize]{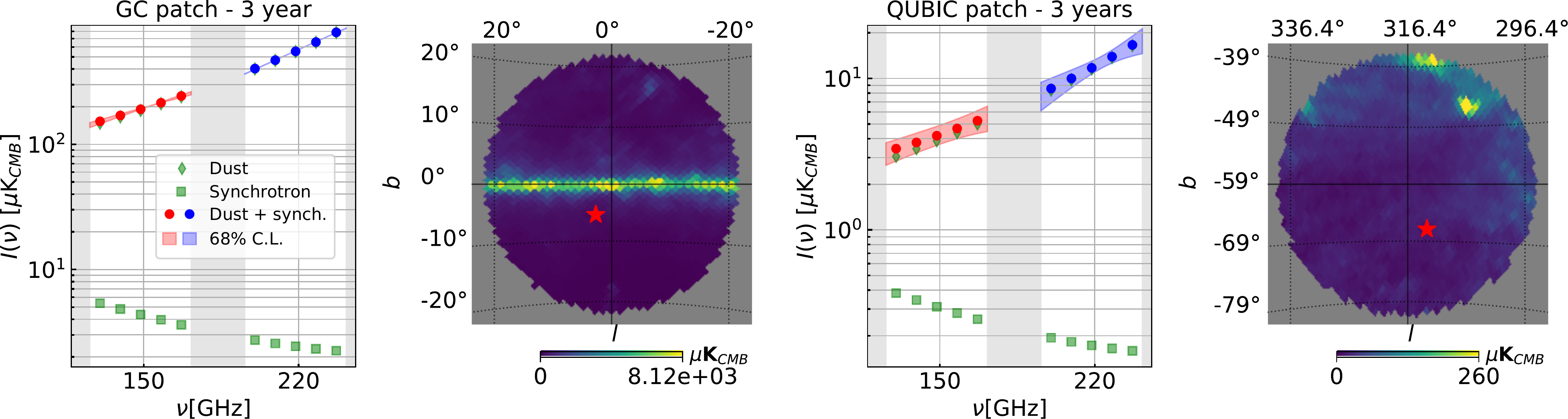}
    \mbox{}\\
    \includegraphics[width=\hsize]{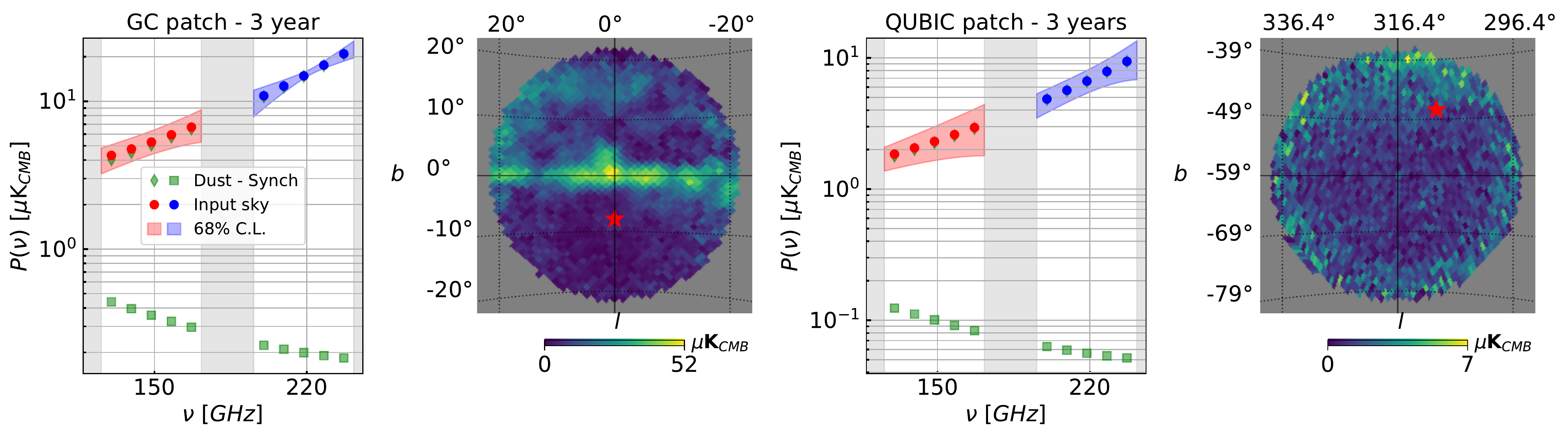}
    \caption{\label{fig:sed-fi} Results of our forecasts on the SED measurement with the QUBIC FI using spectro-imaging with five sub-bands performed independently in both of our wide physical bands (150 and 220~GHz in red and blue respectively). The grey regions corresponds to the unobserved frequencies outside our physical bands. The top row corresponds to total intensity while the bottom one to polarization (with the same fitting model as in equation~\ref{modifiedBB}). Observations with one year sky integration centered on the Galactic center are shown on the left column while the right column shows results in the QUBIC patch [0,-57 deg] (Galactic Coordinates) with three years integration. The red points in the SED are the input convolved to our resolution while the 68\% C.L. regions for our forecasted measurements are shown in light colors. In each case, the corresponding pixel in the reconstructed maps is shown as a red mark.
    We have used $N_\mathrm{side} = 64$ corresponding to $\sim 1$ degree pixels. The maps have a grid spacing in latitude and longitude equal to 10 and 20 degrees respectively.}
    \end{center}
        
\end{figure}

    We now perform the same exercise as in the previous section, but with the expected sensitivity of the QUBIC FI and two focal planes (at 150 and 220~GHz). We explore our ability to constrain the dust SED independently in our two wide physical bands. In figure~\ref{fig:sed-fi} we show the results in total intensity and polarization for one year sky integration toward the Galactic center and three years sky integration on the QUBIC clean patch centered on ($\mathrm{RA}=0$ and $\mathrm{DEC}=-57$). 

These results show that with QUBIC FI and spectro-imaging we will be able to measure the SED of the sky emission in our two bands independently, for both total intensity and polarization, not only near the galactic plane, but in the B-mode search patch itself. Such an approach will allow for more robust constraints of the actual shape of the foregrounds SED, including possible variations within a single wide-band of the dust spectral index without relying on extrapolations between distant frequencies. 

To assess the full potential of spectral imaging for foregrounds control we need to include it into a proper component separation pipeline. This step is the subject of current work and we will report its results in a forthcoming paper.

\subsubsection{Ability to recognize the presence of dust residuals with spectro-imaging}
\label{sss:dust_residuals}

    In this section, we assume that the component separation has already been done and we show that spectro-imaging gives the ability to detect the presence of dust residuals at the tensor-to-scalar likelihood level. Here, we have considered the FI after 3~years of sky integration in the 220~GHz wide band split into 2~sub-bands.

Using the library PySM3~\cite{thorne_python_2017}, we simulate 2~sub-band maps containing dust residuals made with a 0.7~\% fraction of a dust map (d1 PySM model) added to a pure CMB sky with $r=0$. We want to underline here that the value 0.7~\% does not have any other particular meaning other than that of a small fraction of dust residual that leads to the detection of a spurious $r$ of the order of 0.05. This test shows that QUBIC is able to detect it in the data, although a more thorough analysis would involve including the component separation in the analysis, which is outside the scope of this paper.

From those 2~maps, we compute the 3~BB Inter-Band Cross Spectra (IBCS) at effective frequencies $\nu_\mathrm{eff} = \sqrt{\nu_\mathrm{i} \nu_\mathrm{j}}$ where $\nu_\mathrm{i}$ and $\nu_\mathrm{j}$ are the central frequencies of the 2~sub-bands. Then, we perform a likelihood to estimate the tensor to scalar ratio $r$ assuming a pure CMB model (no component separation). The error included in the likelihood estimation is the full multipole-space covariance matrix between the 3~BB IBCS and the bins in $l$ of the spectra obtained with Monte-Carlo simulations. As the sky also contains dust residuals, the likelihood is biased leading to a detection of non-zero tensor-modes which we called $r_\mathrm{dust}$. 

\begin{figure}[t]
    \centering
    \includegraphics[scale = 0.5]{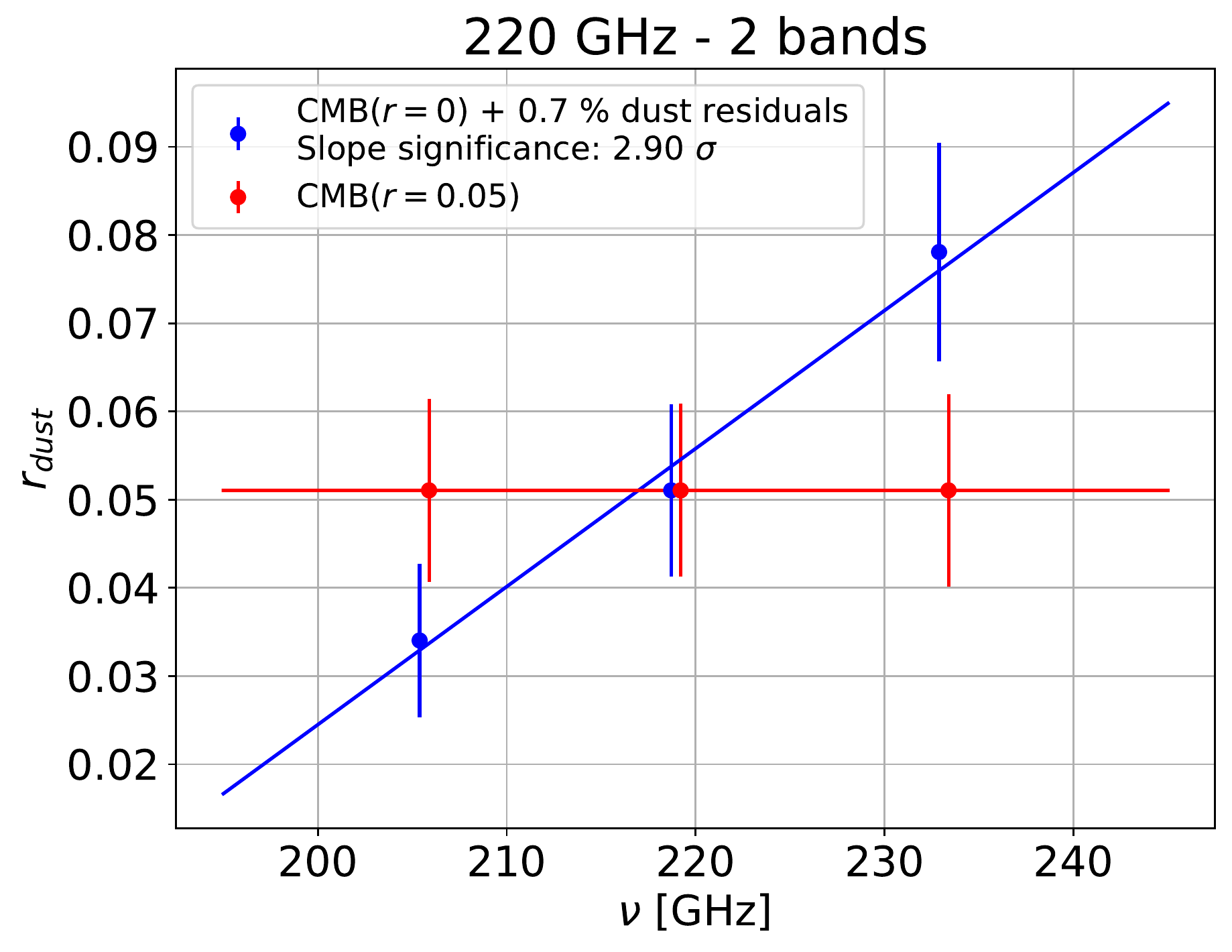}
    \caption{Constraints on the tensor-to-scalar ratio $r$, from a Likelihood analysis, for a sky containing CMB and dust residuals. We consider the 220 GHz wide band split into two sub-bands. This leads to three Inter-Band Cross Spectra (IBCS) at  effective frequencies $\nu_{\mathrm{eff}} = \sqrt{\nu_\mathrm{i} \nu_\mathrm{j}}$ where $\nu_\mathrm{i}$ and $\nu_\mathrm{j}$ are the central frequencies of the two sub-bands. We can distinguish between a pure CMB sky with $r=0.05$ and a sky with CMB ($r=0$) but having 0.7~\% dust residuals. In the first case (red), the measurement of $r_{\mathrm{dust}}$ is constant with the frequency while the presence of dust leads to a slope (blue) which we detect at the $2.9\sigma$ level. In that case, the likelihood estimation was performed on each IBCS separately using the full multipole-space covariance matrix (in contrast with Fig.~\ref{Fig:bmodes} where we have only used the diagonal in order to remain conservative). 
    }
    \label{fig:dust_residuals}
\end{figure}

Instead of having a global estimation of $r_\mathrm{dust}$ over all IBCSs, we can compute a likelihood for each IBCS separately. This gives an estimate of $r_\mathrm{dust}$ for each effective frequency. As the dust emits like a modified black body spectrum, increasing with frequency, the $r_\mathrm{dust}$ estimate increases with frequency. On the contrary, for pure primordial B-modes, the estimated tensor-to-scalar ratio does not depend on frequency. Measuring the evolution of the estimated $r_\mathrm{dust}$ as a function of frequency within the band using spectro-imaging is therefore a powerful tool to estimate our dust contamination independently of any dust modeling.
In Fig.~\ref{fig:dust_residuals}, we show an example where we are able to distinguish the case of pure primordial tensor modes with $r=0.05$ from a sky with no primordial tensor-modes ($r=0$) but 0.7~\% dust residuals. Indeed, in the first case, the measurement of $r_\mathrm{dust}$ is constant with the frequency while the presence of dust leads to a slope that is detected at the 2.9$\sigma$ level.

\section{Conclusions}
\label{sec:conclusions}
In this article, first of a series of eight, we have given an overview of the QUBIC instrument. QUBIC is the first CMB polarimeter using a new technology called ``Bolometric Interferometry'' that combines the background limited sensitivity of bolometric detectors (transition edge sensors) with the clean measurement of interference fringes. QUBIC has been designed to observe the large-scale B-modes of the CMB to detect the elusive tensor perturbations (primordial gravitational waves) created during inflation. A first version of the instrument, with reduced number of horns (64) and detectors (248) operating at 150~GHz, the QUBIC Technological Demonstrator (TD), will be deployed in 2021 and is intended to demonstrate on-sky the capabilities of Bolometric Interferometry. The TD will subsequently be upgraded to the Full Instrument (FI) with the nominal 400~horn array and 992~detectors in each of the two focal planes operating at 150 and 220~GHz.

We have described the general design of QUBIC and have shown that it can be considered as a classical imager that would scan the sky with a beam composed of multiple peaks. These are separated by an angular distance given by the distance between two apertures in the interferometer horn array (in wavelength units), while the resolution of the peaks is given by the maximum size of the interferometer horn array (in wavelength units). We have emphasized two main features of QUBIC. First, the possibility of performing self-calibration~\cite{2013A&A...550A..59B}, similarly as in a classical interferometer, allows us to control instrumental systematic effects. Second,  a bolometric interferometer has simultaneous spatial and spectral sensitivity, thanks to the spectral dependence of the synthesized beam within the physical band of the instrument. This is what allows us to carry out spectral imaging~\cite{2020.QUBIC.PAPER2}.

We have used an extensive set of simulations to make forecasts of the QUBIC performance for both the primordial B-mode search and for testing foregrounds using spectral imaging. These preliminary forecasts assume the detector noise measured during the QUBIC calibration campaign~\cite{2020.QUBIC.PAPER3} and a stable atmosphere corresponding to the QUBIC site in Argentina at 5000\,m a.s.l. In this work we did not consider neither component separation nor the effect of fluctuations in the atmospheric load, both of which require a significant step forward in our pipeline that we are currently carrying out. However, we showed that regardless of component separation spectral imaging allows us to reconstruct the sky emission SED pixel-by-pixel and to recognize the presence of foregrounds residuals in the derived tensor-to-scalar ratio.

These simulations show that the QUBIC TD will be able to demonstrate spectral imaging with one year of sky integration on a field centered on the Galactic center. The QUBIC TD will measure the SED of bright regions in both intensity and polarization. After the upgrade to the FI, QUBIC will reach its nominal configuration allowing us to reach a sensitivity to B-modes corresponding to a 68\% C.L. upper-limit on the effective tensor-to-scalar ratio (primordial tensors + dust) $\sigma(r)=0.015$ with three years of sky integration. We have also shown how QUBIC FI will be able to put constraints on the dust contamination through direct measurement of the dust SED and properties in the reconstructed maps at multiple sub-frequencies within a physical band, or through a study of the evolution of the recovered tensor-to-scalar ratio in sub-bands as a function of frequency.

Future studies will explore in detail the impact of atmosphere fluctuations and how spectral imaging can lead to improved component separation using classical techniques as well as techniques specific to bolometric interferometry.

\appendix

\section{Impact of atmospheric fluctuations on scales smaller than the synthesized beam cutoff angles.}
\label{app_impact_small_scale_atm}

    In section~\ref{sec_map_making_noise_structure} we have seen that the synthesized beam naturally filters-out large-scale fluctuations from atmospheric gradients. We have also explained that to understand the effect of atmosphere fluctuations on scales smaller than the synthesized beam cutoff angles (8.8\,degrees at 150\,GHz and 6\,degrees at 220\,GHz) it is necessary to build simulations based on atmosphere measurements taken at the site. Here we want to show that the available data taken at sites similar to that of QUBIC allows us to say that the instrument will be sensitive to fluctuations in a defined range of turbulence scales.

    Fluctuations in the atmospheric radiation load are determined mainly by turbulent cells in the water vapor column that are distributed over scales within an interval $\left[L_0^\mathrm{min}, L_0^\mathrm{max}\right]$. The values of the minimum and maximum cutoff coherence scales, $L_0^\mathrm{min}$ and $L_0^\mathrm{max}$, depend on the physical properties of the atmosphere and are typical of each observations site.
    
    In a study conducted in the framework of the POLARBEAR experiment Errard et al.~\cite{errard2015} studied the distribution of various atmosphere parameters at Atacama and showed that turbulence scales are distributed in the interval $L_0 \in \left[200\,\mathrm{m}, 800\,\mathrm{m}\right]$ with a peak around 300\,m (see the left panel of figure~11 in Errard et al.~\cite{errard2015}). Assuming that the atmosphere at Alto Chorrillo is not very different from that at Atacama, we can use this distribution to assess quickly the impact of atmosphere fluctuations on scales smaller than the synthetic beam cutoff angles (8.8~degrees at 150\,GHz and 6~degrees at 220\,GHz).
    
    Let us consider a fluctuations correlation length, $L_0$. This length corresponds to angular scales $\theta^\mathrm{atm}_\mathrm{corr}(z)\sim L_0/z$, where $z$ is the height above the telescope line-of-sight.

    \begin{figure}[!t]
        \begin{center}
            \includegraphics[width=14cm]{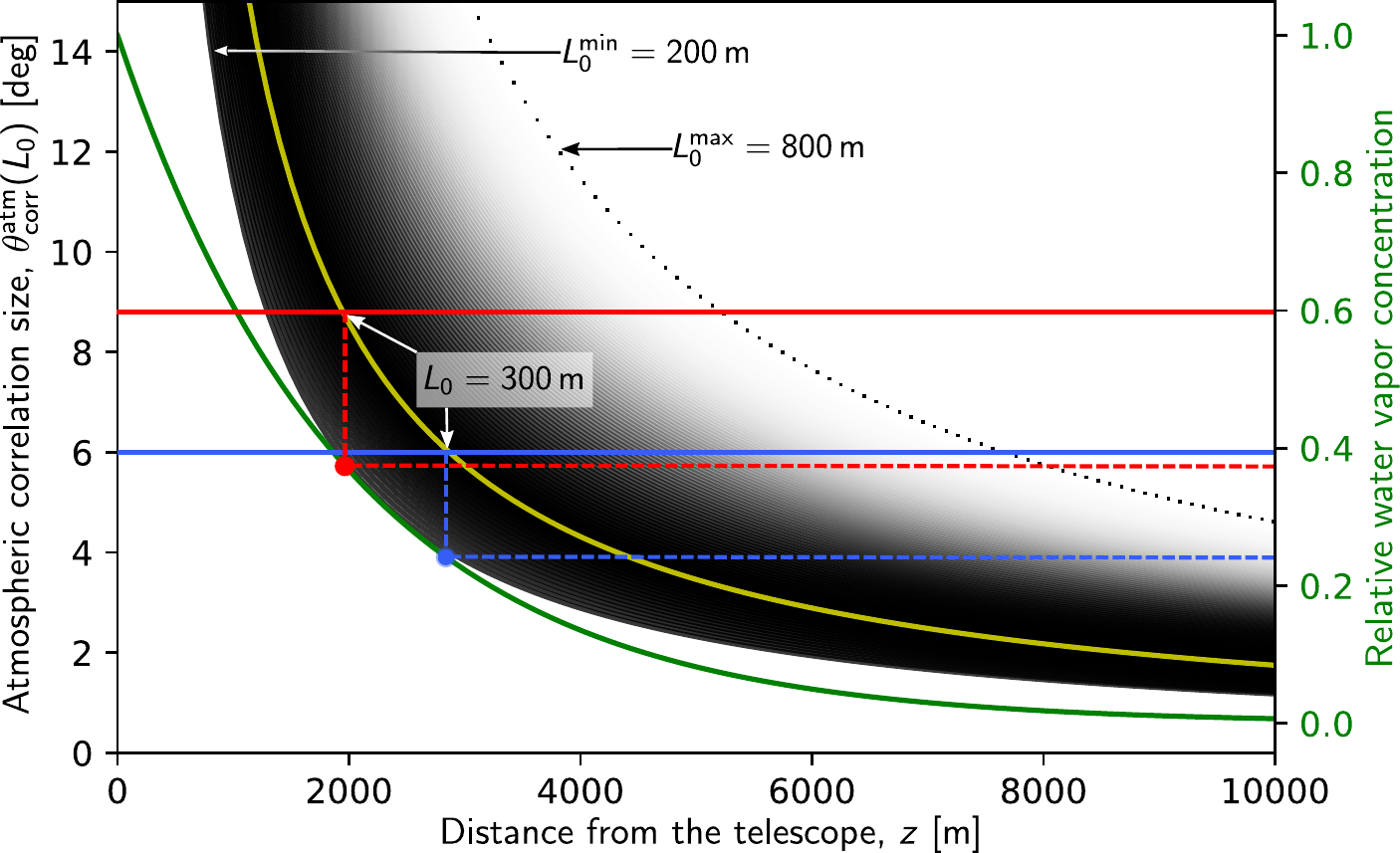}
            \caption{\label{fig_small_scale_atmo}Relationship between the relative water vapor concentration and atmospheric turbulence correlations angular scales. Black lines are a family of curves $\theta^\mathrm{atm}_\mathrm{corr}(z)$ corresponding to values of $L_0$ in the interval $\left[200\,\mathrm{m}, 800\,\mathrm{m}\right]$. The curves are marked with a transparency level that reflects the distribution in figure~11 of Errard et al.~\cite{errard2015}. The yellow curve marks the case with $L_0^\mathrm{max}=300\,\mathrm{m}$, corresponding to the most likely correlation scale. The red and blue horizontal lines mark the synthetic beam cutoff angles at 150 and 220\,GHz. The green curve (right axis) reports the relative water vapor concentration versus height, assuming a height scale of 2000\,m. The dashed red and blue lines are guides that show the water vapor concentration corresponding to the synthesized beam cutoff scales and the case $L_0^\mathrm{max}$. See text for a more detailed description.}
        \end{center}
    \end{figure}
    
    In figure~\ref{fig_small_scale_atmo} we show, in black, a family of curves $\theta^\mathrm{atm}_\mathrm{corr}(z)$ corresponding to values of $L_0$ in the interval $\left[200\,\mathrm{m}, 800\,\mathrm{m}\right]$. The curves are marked with a transparency level that reflects the distribution in figure~11 of Errard et al.~\cite{errard2015}. The yellow curve marks the case with $L_0^\mathrm{max}=300\,\mathrm{m}$, corresponding to the most likely correlation scale. The red and blue horizontal lines mark the synthetic beam cutoff angles at 150 and 220\,GHz, so that for small scale fluctuations we are interested in the regions below these lines. The green curve (right axis) reports the relative water vapor concentration versus height, assuming a height scale of 2000\,m\footnote{The water vapor scale height is the distance over which a the water vapor content decreases by a factor of $e$}, which can be considered representative for Alto Chorrillo (see, for example, the left panel of figure 1 in Mararieva et al.~\cite{makarieva2013}). Finally, the dashed red and blue lines are guides that show the water vapor concentration corresponding to the synthesized beam cutoff scales and the case $L_0^\mathrm{max}$.
    
    Let us take, for example, the most likely correlation scale, $L_0^\mathrm{max}$. The largest angular scales connected to $L_0^\mathrm{max}$ are the synthetic beam cutoff scales and, with these values, we see that we will be affected by fluctuations involving $\sim$40\% of the water vapor column at 150\,GHz and slightly more than 20\% at 220\,GHz. It is worth noting that the bottom-left and top-right white areas of the plot correspond to angular scales where we do not expect significant fluctuations. For example, for $z = 2000\,\mathrm{m}$ we do not expect fluctuations on angular scales up to 6\,degrees.
    
    From this simple analysis we can conclude that the effect from turbulent cells larger than 300\,m will be increasingly negligible, while we may expect some impact from cells in the range $\left[200\,\mathrm{m}, 300\,\mathrm{m}\right]$. Of course this analysis does not answer yet the question of what effect we can expect on the final science from atmosphere fluctuations on these scales. A complete assessment will be performed on the basis of dedicated atmospheric measurements taken at the QUBIC site.

\section*{Acknowledgements}
\label{sec:ack}
We thank the anonymous referees as well as the editor for insightful reviews that greatly improved this work.
QUBIC is funded by the following agencies. France: ANR (Agence Nationale de la Recherche) 2012 and 2014, DIM-ACAV (Domaine d’Interet Majeur-Astronomie et Conditions d’Apparition de la Vie), Labex UnivEarthS (Université de Paris), CNRS/IN2P3 (Centre National de la Recherche Scientifique/Institut National de Physique Nucléaire et de Physique des Particules), CNRS/INSU (Centre National de la Recherche Scientifique/Institut National des Sciences de l’Univers). Italy: CNR/PNRA (Consiglio Nazionale delle Ricerche/ Programma Nazionale Ricerche in Antartide) until 2016, INFN (Istituto Nazionale di Fisica Nucleare) since 2017. Argentina: MINCyT (Ministerio de Ciencia, Tecnología e Innovación), CNEA (Comisión Nacional de Energía Atómica), CONICET (Consejo Nacional de Investigaciones Científicas y Técnicas).

\afterpage{\clearpage}
\bibliographystyle{ieeetr}
\bibliography{qubic}  

\end{document}